\documentclass[12pt]{iopart}

\usepackage{graphicx}
\usepackage{epsfig,color}

\expandafter\let\csname equation*\endcsname\relax

\expandafter\let\csname endequation*\endcsname\relax

\usepackage{amsmath,amssymb,txfonts}
\usepackage{hyperref}
\hypersetup{
    colorlinks=true,
    citecolor=black,
    filecolor=black,
    linkcolor=black,
    urlcolor=black
}
\usepackage{gensymb}
\usepackage{color}

\graphicspath{ {figures/} }
\usepackage{cite}

\newcommand{\ket}[1]{\left| #1\right\rangle}
\newcommand{\bra}[1]{\left\langle #1 \right|}
\newcommand{\ex}[1]{\langle #1 \rangle}
\begin{document}

\title{Quantum Simulation of the Abelian-Higgs Lattice Gauge Theory with Ultracold Atoms}

%\author{Daniel Gonz\'{a}lez-Cuadra}
%\address{ICFO -- The Institute of Photonic Sciences, Av. C.F. Gauss 3, E-08860, Castelldefels (Barcelona), Spain}
%\address{Max-Planck-Institut f\"{u}r Quantenoptik, Hans-Kopfermann-Straße 1, D-85748 Garching, Germany}
%\author{Erez Zohar}
%\address{Max-Planck-Institut f\"{u}r Quantenoptik, Hans-Kopfermann-Straße 1, D-85748 Garching, Germany}
%\author{J. Ignacio Cirac}
%\address{Max-Planck-Institut f\"{u}r Quantenoptik, Hans-Kopfermann-Straße 1, D-85748 Garching, Germany}

\author{%
Daniel Gonz\'{a}lez-Cuadra$^{1,2}$,
Erez Zohar$^{2}$
and
J. Ignacio Cirac$^{2}$
}

\address{$^{1}$~ICFO -- The Institute of Photonic Sciences, Av. C.F. Gauss 3, E-08860, Castelldefels (Barcelona), Spain}
\address{$^{2}$~Max-Planck-Institut f\"{u}r Quantenoptik, Hans-Kopfermann-Stra\ss e 1, D-85748 Garching, Germany}

\begin{abstract}
We present a quantum simulation scheme for the Abelian-Higgs lattice gauge theory using ultracold bosonic atoms in optical lattices. The model contains both gauge and Higgs scalar fields, and exhibits interesting phases related to confinement and the Higgs mechanism. The model can be simulated by an atomic Hamiltonian, by first mapping the local gauge symmetry to an internal symmetry of the atomic system, the conservation of hyperfine angular momentum in atomic collisions. By including auxiliary bosons in the simulation, we show how the Abelian-Higgs Hamiltonian emerges effectively. We analyze the accuracy of our method in terms of different experimental parameters, as well as the effect of the finite  number of bosons on the quantum simulator. Finally, we propose possible experiments for studying the ground state of the system in different regimes of the theory, and measuring interesting high energy physics phenomena in real time.
\end{abstract}

%\submitto{\NJP}
%\maketitle

\section{Introduction}

The study of natural systems governed by the laws of quantum mechanics is a complicated task. Most of the usual theoretical and numerical techniques employed to tackle the behaviour of quantum systems become rapidly overwhelmed once their size starts increasing. In particular, the computational resources required to simulate a quantum mechanical system using a classical computer scale exponentially with the number of its constituents. This situation led Richard Feynman to think of the concept of a \textit{quantum simulator} \cite{Feynman1982}. According to his idea, such a device would be governed itself by the laws of quantum mechanics and, exploiting this fact, could be used to study other quantum mechanical systems more efficiently. Feynman's idea was formalized as an algorithm that could be processed by an \textit{universal quantum computer} \cite{Lloyd}. This is referred to as a \textit{digital} quantum simulation, since it is based on approximating the simulated system's dynamics by applying a discrete set of operations (quantum gates) on a highly-controllable quantum device. Quantum computers are very interesting from a theoretical point of view, since, in principle, they could simulate any other physical system using only polynomial resources, provided that certain locality conditions are fulfilled \cite{Lloyd}. From the experimental side, however\textemdash and in spite of the great advances over the last decades regarding the manipulation of microscopic quantum systems\textemdash a practical implementation of a fully operational quantum computer is still a long-term goal.

Notwithstanding, having these controllable devices so well studied and reachable makes quantum simulation a reality, by means of simpler quantum devices that can perform calculations beyond the scope of classical machines. In particular, the so-called \textit{analog} quantum simulators, although much more limited than an universal simulator, can be used to study a broad range of quantum systems within the reach of today's technological progress \cite{Cirac_goals,Buluta_Nori}. Analog simulators act by mapping the degrees of freedom of the simulated system to those of the simulating one. The latter can be controlled in the laboratory and its dynamics can be tailored (in particular, the corresponding Hamiltonian) to be equivalent to those of the system we are trying to study. This allows us to obtain information about systems that can not be accessed experimentally, by investigating others for which state preparation and measurements are much easier tasks.

Among the relevant platforms that can serve as quantum simulators, both analog and digital, many atomic and optical systems stand out due to their remarkable experimental controllability. Some examples include ultracold atoms \cite{Jaksch,toolbox,Bloch,Lewenstein_paper,Lewenstein}, trapped ions \cite{Cirac_ions,Wineland,Blatt_Roos,Zoller,Porras}, photonic systems \cite{photonic} and Rydberg atoms \cite{rydberg}. Ultracold atoms in optical lattices, in particular, present the possibility of recreating many different interactions\textemdash such as nearest-neighbor, long-range forces, on-site interactions, etc\textemdash allowing for the simulation of both condensed matter and high energy physics models. Solid-state systems, such as quantum dots \cite{DiVincenzo,quantum_dot_1,quantum_dot_2} or superconducting circuits \cite{superconducting_1,superconducting_2}, also show prominent results that make them interesting candidates to perform quantum simulations.

Using these ideas, many condensed matter Hamiltonians have been considered for quantum simulations, some of them even realized experimentally. Some examples include spin systems \cite{Porras,Porras_2,Ripoll_1,Ripoll_2}, such as the Ising or Heisenberg models; the Bose-Hubbard model \cite{Jaksch,Greiner}; the Tonks-Girardeau gas \cite{Paredes}, or copper-oxide superconductors \cite{quantum_dot_1}. External gauge potentials can be simulated as well, allowing for the study of the fractional quantum Hall effect \cite{fractional_quantum_hall} and other topological phenomena \cite{Hofstadter,escher_staircase,non_abelian_external_gauge_1,non_abelian_external_gauge_2}.

Quantum simulations of high energy physics models, although more demanding than their condensed matter counterparts, are also possible. Some examples include the simulation of the Dirac \cite{Dirac_1,Dirac_2,Dirac_3} and Majorana equations \cite{Majorana}, the Casimir force \cite{Casimir}, the Schwinger mechanism \cite{Schwinger_mechanism} or the oscillations of neutrinos \cite{neutrino,neutrino_2}. Simulations of quantum field theories \cite{Preskill,relativistic_field_theory}, gravitational theories \cite{gravity} and black holes \cite{black_holes_1,black_holes_2} have been proposed in the last years, some of them realized experimentally as well \cite{black_holes_3}.

Within high energy physics, gauge theories are particularly relevant, since they lie in the core of the Standard Model of particle physics \cite{Peskin,Schwartz,Polyakov,Ryder}: dynamical gauge degrees of freedom are introduced in quantum field theories to explain the interactions between the basic constituents of matter, such as quarks or electrons. Many techniques have been developed to study such theories. Some of them rely on perturbative expansions around small coupling constants. However, these methods lose their validity if one tries to apply them for the study of \textit{non-perturbative} phenomena, where the relevant coupling constant presents large values. This is the case for the interaction strength between separated quarks, which grows with the distance between them (running coupling) \cite{Peskin,Schwartz}. At short distances, the interaction's coupling constant presents small values (asymptotic freedom) \cite{Wilczek}, and the perturbative methods can be applied. At long distances, however, the growth in the interaction strength gives rise to the \textit{confinement} mechanism, implying that no free quarks can be found in nature, a claim that is supported by the experiments \cite{Schwartz}. To deal with this and other non-perturbative phenomena, the lattice formulation of gauge theories was developed \cite{Wilson,Kogut}, obtaining a proper framework to perform numerical simulations on these theories\textemdash using Monte Carlo methods, in particular \cite{Smit}.

These techniques provided a great advance in the understanding of particle physics during the last decades. However, they present some limitations when they are applied to certain cases. An example is the \textit{sign problem}, which appears in regimes with a finite chemical potential for fermionic particles \cite{sign_problem}, and becomes problematic for the studying of different phases of QCD, such as the quark-gluon plasma or the color-superconducting phase \cite{phases_QCD}. The analysis of real-time dynamics is also lacking, since Monte Carlo simulations only allow to calculate Euclidean space-time correlation functions in imaginary time (after a Wick rotation).

In this context, two different approaches have been recently proposed to study lattice gauge theories in regimes that can not be accessed using previous techniques. One involves the application of methods based on tensor networks \cite{TN_1,TN_3,TN_4,TN_5,TN_6,TN_7,TN_8,TN_9,TN_10,TN_11,TN_12,TN_13,TN_14,TN_15}. Another consists on performing quantum simulations using low energy quantum systems. In contrast to the case of condensed matter models, lattice gauge theories are more complicated to simulate using low energy systems, such as ultracold atoms, since the gauge and Lorentz symmetries are not naturally present in the latter. In addition, simulating both gauge and matter fields usually requires the use of bosonic and fermionic atoms, which increases the experimental complexity. Any attempt to simulate theses theories must take these facts into account \cite{Erez_5,Erez_review}.

Some examples of simulation proposals for gauge theories using ultracold atoms include both the continuous \cite{continuous_QED} and lattice version of QED, focusing especially on the latter case. In particular, analog simulations for cQED were proposed both in the absence \cite{Erez_1,Erez_2} and presence of dynamical matter \cite{QED_matter,Erez_3,123,luca}, where gauge invariance emerges as an effective symmetry. Realizations of $U(1)$ gauge theories with dynamical or background Higgs fields, using effective gauge invariance, were discussed in \cite{abelian_higgs_1,abelian_higgs_2,abelian_higgs_3}. In \cite{Erez_5,kasper_2,Kasper} the gauge symmetry is obtained exactly by mapping it to an internal symmetry of the atomic system. Other quantum systems that have been considered for quantum simulations of lattice gauge theories are superconducting circuits \cite{134,135}, trapped ions \cite{ions_gauge,Hauke} and Rydberg atoms  \cite{rydberg_gauge,rydberg}. Discrete symmetry groups have been considered as well, both for analog \cite{discrete_gauge, Erez_5,discrete_gauge_2} and digital simulations \cite{digital_gauge_1,digital_gauge_2}, as well as non-abelian theories \cite{Erez_4,non_abelian_gauge_1,non_abelian_gauge_2,non_abelian_digital,151,152,153,154}. A general formalism for digital simulations of lattice gauge theories with ultracold atoms can be found in \cite{digital_gauge_1}.

Recently, the first experimental realization of a quantum simulation of a lattice gauge theory was performed \cite{simulation_experiment}, where the real-time dynamics of the Schwinger model were simulated on a few-qubit trapped-ion quantum computer. This experiment opened the door for studying high energy physics beyond the capabilities of classical simulations.

The present work is devoted to the quantum simulation of the Abelian-Higgs lattice gauge theory, which contains both gauge and Higgs scalar fields, using ultracold bosonic atoms confined in optical lattices. This will be done by mapping atomic symmetries to local gauge invariance. The paper is organized as follows. In Section \ref{sec:the_model}, we introduce the Abelian-Higgs model, which is used as toy model in high energy physics to study the Higgs mechanism. This model admits a discretized description as a lattice gauge theory, whose phase diagram shows, apart from the mentioned mechanism, a confinement phase for the scalar Higgs field. We consider the Hamiltonian formulation of the model, more suited for quantum simulation purposes. In Section \ref{sec:quantum_simulation}, we introduce the necessary ingredients to simulate the Abelian-Higgs Hamiltonian using ultracold atoms in optical lattices. First, the basic properties of the simulating system are explained, starting with the second-quantized Hamiltonian that describe an atomic system in the ultracold regime, as well as the most important experimental techniques that are needed to control and manipulate it. Then, we show how the degrees of freedom and interactions of the simulated model\textemdash and, in particular, the local gauge symmetry\textemdash can be mapped onto the atomic system. With the help of auxiliary ``hard-core" bosons, we show how the complete Abelian-Higgs Hamiltonian emerges effectively up to some small corrections. Finally, in Section \ref{sec:possible_experiments}, we introduce possible experiments to measure interesting high energy physics phenomena, such as the dynamical breaking of electric flux lines between scalar charges, effect that is related to confinement \cite{confinement_problem}.

\section{The Simulated Model}
\label{sec:the_model}

The spontaneous breaking of a symmetry refers to a situation characterized by a ground state which is not invariant under the same set of transformations as the theory's Lagrangian or Hamiltonian. The Higgs mechanism \cite{Higgs_Anderson,Higgs_Englert,Higgs_Higgs,Higgs_Kibble} describes this phenomenon in the presence of a (local) gauge symmetry. It can be applied, for instance, to the $S\!U(2)\times U(1)$ gauge group, associated to the electroweak interaction in the Standard Model of particle physics, explaining how the corresponding gauge bosons, $W^\pm$ and $Z$, acquire a non-zero mass \cite{Weinberg}. The basic features of the mechanism can be studied, however, in a much simpler case, the \textit{Abelian-Higgs model} \cite{Peskin,abelian_higgs_5}, involving complex scalar fields coupled to abelian gauge fields (\ref{app:higgs}).

The Abelian-Higgs model can be formulated as a lattice gauge theory \cite{Fradkin_Shenker,Einhorn_Savit_1,Einhorn_Savit_2,Jones_Kogut_Sinclair,Ostewalder_Seiler,Abelian_Higgs_Montecarlo,abelian_higgs_4}, this is, as a field theory on a discretized space-time, invariant under $U(1)$ local transformations. The phase diagram of the theory shows very interesting phenomena, related to the Higgs mechanism and the confinement phenomenon (\ref{app:action}).

For quantum simulation purposes, it is more convenient to express the theory in a Hamiltonian formulation. This makes the mapping between the simulating and the simulated system\textemdash in this case, ultracold atoms in optical lattices\textemdash more straightforward, since the latter is described usually in terms of a second-quantized many-body Hamiltonian. Here, we shall introduce the Abelian-Higgs model directly in the Hamiltonian formulation. We obtained the former by applying the transfer matrix method to the standard lattice gauge action used in the high energy physics literature. The details of these calculations can be found in \ref{sec:hamiltonian_formulation}.

The Abelian-Higgs theory involves two types of fields: gauge and Higgs fields. We will formulate it on a $d$ dimensional spatial square lattice, with continuous time, and will fix the lattice spacing $a=1$. 

The gauge fields reside on the links of the lattice. They are represented by the unitary operators $\hat{U}^{(q)}_{\mathbf{n},k}\equiv e^{-iq\hat{\theta}_{\mathbf{n},k}}$, residing on the links (joining the vertices $\mathbf{n}$ and $\mathbf{n}+\,\hat{k}$, with $\hat{k}$ a unit vector in one of the $d$ orthogonal directions). $q$ is an integer denoting the $U(1)$ representation; We will omit it from now on and focus on the case $q=1$. On each link we introduce another operator, $\hat{E}_{\mathbf{n},k}$, with the commutation relations
\begin{equation}
\label{eq:commutator_U_E}
[\hat{E}^{\vphantom{\dagger}}_{\mathbf{n},k},U^\dagger_{\mathbf{n}^\prime,k^\prime}]=\delta^{\vphantom{\dagger}}_{k,k^\prime}\delta^{\vphantom{\dagger}}_{\mathbf{n},\mathbf{n}^\prime}U^\dagger_{\mathbf{n},k}\,.
\end{equation}
$\hat{U}^{\vphantom{\dagger}}_{\mathbf{n},k}$ and $\hat{U}^\dagger_{\mathbf{n},k}$ are unitary operators acting as ladder operators with respect to $\hat{E}_{\mathbf{n},k}$, which has a discrete spectrum,
\begin{equation}
\begin{aligned}
\hat{E}_{\mathbf{n},k}\ket{m}_{\mathbf{n},k}&=m_{\mathbf{n},k}\ket{m}_{\mathbf{n},k}, & m_{\mathbf{n},k}&\in\mathbb{Z}\,.
\end{aligned}
\end{equation}
We will call the latter the \textit{electric field} operator. The commutation relations (\ref{eq:commutator_U_E}) define how the operators $\hat{U}^{\vphantom{\dagger}}_{\mathbf{n}\vphantom{\hat{k}},k}$ and $\hat{U}^\dagger_{\mathbf{n}\vphantom{\hat{k}},k}$ act on the eigenstates of $\hat{E}_{\mathbf{n}\vphantom{\hat{k}},k}$,
\begin{equation}
\begin{aligned}
\hat{U}_{\mathbf{n}\vphantom{\hat{k}},k}\ket{m}_{\mathbf{n},k}&=\ket{m-1}_{\mathbf{n},k}, & \hat{U}^\dagger_{\mathbf{n},k}\ket{m}_{\mathbf{n},k}&=\ket{m+1}_{\mathbf{n},k}.
\end{aligned}
\end{equation}

The Higgs fields, $\hat{\phi}_{\mathbf{n}}\equiv e^{-i\hat{\varphi}_{\mathbf{n}}}$, which reside on the vertices of the lattice, are unitary operators as well.
 Their Hilbert spaces are very similar to these of the gauge fields; we define the operator
 $\hat{Q}_{\mathbf{n}}$\textemdash referred as the \textit{charge} operator\textemdash obeying the commutation relations
\begin{equation}
\label{eq:commutation_phi_Q}
[\hat{Q}^{\vphantom{\dagger}}_\mathbf{n},\hat{\phi}^\dagger_{\mathbf{n}^\prime}]=\delta^{\vphantom{\dagger}}_{\mathbf{n},\mathbf{n}^\prime}\hat{\phi}^\dagger_\mathbf{n}\,,
\end{equation}
having $\hat{Q}_{\mathbf{n}}$, again, a discrete spectrum. $\hat{\phi}^\dagger_\mathbf{n}$ and $\hat{\phi}^{\vphantom{\dagger}}_\mathbf{n}$ are unitary operators that act as raising and lowering operators, respectively,
\begin{equation}
\begin{aligned}
\hat{\phi}_{\mathbf{n}\vphantom{\hat{k}}}\ket{Q}_{\mathbf{n}}&=\ket{Q-1}_{\mathbf{n}}, & \hat{\phi}^\dagger_{\mathbf{n}}\ket{Q}_{\mathbf{n}}&=\ket{Q+1}_{\mathbf{n}},
\end{aligned}
\end{equation}
where $\ket{Q}_{\mathbf{n}}$ is an eigenstate of the charge operator $\hat{Q}_{\mathbf{n}}$.

\begin{figure}[t]
  			\centering
    		\includegraphics[width=0.3\textwidth]{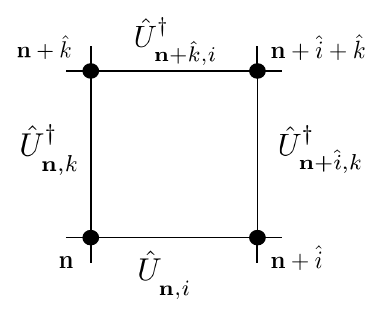}
    		\caption{The plaquette term contains the product of four gauge-field terms corresponding to the four links of a plaquette.}
    		\label{fig:plaquette}
\end{figure}

All these ingredients appear in the \textit{Abelian-Higgs Hamiltonian},
%In order to achieve this, the transfer matrix method is applied to the action (\ref{eq:lattice_gauge_action}). This method was first developed in statistical physics \cite{Schultz}, and was later applied to obtain the Hamiltonian that describes lattice gauge theories \cite{Kogut}\textemdash first studied by Kogut and Susskind \cite{Kogut-Susskind,lattice_fermions}. Here, the transfer matrix method is applied for scalar matter corresponding the Higgs field transforming under the fundamental representation of the $U(1)$ gauge group ($q=1$), rendering, upon taking the continuum limit in the time direction, a quantum Hamiltonian defined on a spatial lattice. The details of this derivation can be found in \ref{app:transfer_matrix}, resulting in the following \textit{Abelian-Higgs Hamiltonian},
\begin{equation}
\label{eq:my_Hamiltonian}
\begin{aligned}
&\hat{H}=\frac{1}{2R^2}\sum_\mathbf{n}\hat{Q}^2_\mathbf{n}-\frac{R^2}{2}\sum_{\mathbf{n},k}\left[\hat{\phi}^\dagger_{\mathbf{n}\vphantom{\hat{k}}}\hat{U}^\dagger_{\mathbf{n}\vphantom{\hat{k}},k}\hat{\phi}^{\vphantom{\dagger}}_{\mathbf{n}+\hat{k}}+h.c.\right]\\
&+\frac{g^2}{2}\sum_{\mathbf{n},k}\hat{E}^2_{\mathbf{n},k}-\frac{1}{2g^2}\sum_{\mathbf{n},ik}\left(\hat{U}^{\vphantom{\dagger}}_{\mathbf{n}\vphantom{+\hat{k}},i}\hat{U}^{\vphantom{\dagger}}_{\mathbf{n}+\hat{i},k}\hat{U}^\dagger_{\mathbf{n}+\hat{k},i}\hat{U}^\dagger_{\mathbf{n}\vphantom{+\hat{k}},k}+h.c.\right),
\end{aligned}
\end{equation}
where the sums in $\mathbf{n}$ and $k$ run over the vertices and links of a spatial lattice in $d$ dimensions, and the coupling constants  $R$ and $g$ are introduced to recover the continuous Abelian-Higgs model in the limit $a\rightarrow 0$.  The Hamiltonian (\ref{eq:my_Hamiltonian}) is composed of a non-interacting part with local terms (charge and electric field) for the vertices and links of the lattice, and an interacting part that includes a plaquette-type term (Fig. \ref{fig:plaquette}) that creates electric field excitations along the four links of a plaquette, and nearest-neighbor vertex couplings, mediated by an excitation on the link that joins them. The second row in equation (\ref{eq:my_Hamiltonian}) corresponds, in particular, to the pure-gauge Kogut-Susskind Hamiltonian \cite{Kogut-Susskind}.

%\subsection{Gauss's Law}
%\label{sec:gauss_law}

%Before studying the ground state of the Abelian-Higgs Hamiltonian (\ref{eq:my_Hamiltonian}), let us explain the role that gauge symmetry plays in this formulation.

In the Hamiltonian formulation of lattice gauge theories, the gauge  is fixed in the temporal direction, $\theta_{n,0}=0$ (\ref{sec:hamiltonian_formulation}). However, it is not fixed in the spatial directions. Therefore, the Hamiltonian is still gauge invariant under a restricted subset of local transformations applied to the vertices of the spatial lattice,
\begin{equation}
\label{eq:gauge_transformation_vertices}
V_\mathbf{n}=e^{i\alpha_\mathbf{n}}.
\end{equation}

In this formalism, the gauge transformations are generated, like any other continuous symmetry, by operators $\hat{G}_\mathbf{n}$ that commute with the Hamiltonian of the theory,
\begin{equation}
\label{eq:commutator_H_G}
[\hat{H},\hat{G}_\mathbf{n}]=0\,,
\end{equation}
defined, in this case, locally for each vertex. In the case of the Abelian-Higgs Hamiltonian (\ref{eq:my_Hamiltonian}), the generators $\hat{G}_\mathbf{n}$ are expressed as
\begin{equation}
\label{eq:gauss_law}
\hat{G}_{\textbf{n}\vphantom{\hat{k}}}=\sum_k\left(\hat{E}_{\textbf{n}\vphantom{\hat{k}},k}-\hat{E}_{\textbf{n}-\hat{\textbf{k}},k}\right)-\hat{Q}_{\mathbf{n}\vphantom{\hat{k}}}\,.
\end{equation}
It is easy to check that such operator commutes with the Hamiltonian for every vertex $\mathbf{n}$.

The eigenvalues of the operators $\hat{G}_\mathbf{n}$ are called static charges $q_\mathbf{n}$, taking, as well, integer values. The Hilbert space $\mathcal{H}$ of the system is divided into sectors, each one associated to a different static charge configuration,
\begin{equation}
\label{eq:sectors}
\mathcal{H}=\bigoplus_{\{q_\mathbf{n}\}}\mathcal{H}(\{q_\mathbf{n}\})\,,
\end{equation}
such that
\begin{equation}
\label{eq:eigen_G}
\begin{aligned}
\hat{G}_\mathbf{n}\ket{\psi(\{q_\mathbf{n}\})}&=q_\mathbf{n}\ket{\psi(\{q_\mathbf{n}\})}, & &\forall\,\ket{\psi(\{q_\mathbf{n}\})}\in\mathcal{H}(\{q_\mathbf{n}\})\,.
\end{aligned}
\end{equation}

Since the Hamiltonian commutes with the generators of the gauge transformations, the dynamics can not mix different sectors. If the initial state  belongs to a certain sector, the system will remain in the same sector through its time evolution. The above equation is known as the \textit{Gauss's law}, and it is crucial for the structure of the physical states of the system.

\begin{figure}[t]
  			\centering
    		\includegraphics[width=0.3\textwidth]{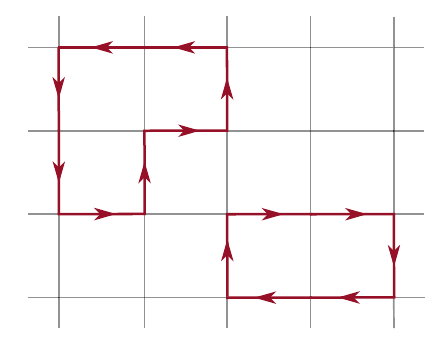}
    		\caption{The pure-gauge excitations on top of the empty state consist of closed electric flux lines, created by the plaquette interaction term in the Hamiltonian (\ref{eq:my_Hamiltonian}). Since the plaquette term is gauge-invariant, these states are gauge invariant as well. This is clear observing that the Gauss law (\ref{eq:gauss_law}) is fulfilled on each vertex, forcing the closed loop structure. In the figure, positive values of the electric field are represented with arrows pointing right or upwards.}
    		\label{fig:discrete_wilson_loop}
\end{figure}

In general, the total Hilbert space of the system would be the product over the vertices and links of the lattice of each infinite dimensional local Hilbert space. However, Gauss law (\ref{eq:gauss_law}) imposes a constraint on the allowed states where the Hamiltonian (\ref{eq:my_Hamiltonian}) acts. Let us first consider the pure-gauge Kogut-Susskind Hamiltonian, for the case $R=0$, within the zero-static charge sector ($q_\mathbf{n}=0$ $\forall \mathbf{n}$). In the strong coupling limit ($g\gg 1$), the ground state of the system is the empty state. This is, every link of the lattice is an eigenstate of the electric field operator with eigenvalue equal to zero. As the value of the coupling constant $g$ decreases, the empty state is no longer the ground state of the system. The plaquette-interaction term can create excitations composed of closed electric flux lines. The ground state, and the rest of excited states, will be superposition of such loops, with an increasing number of them when the energy of the system grows (Fig. \ref{fig:discrete_wilson_loop}).

\begin{figure}[t]
  			\centering
    		\includegraphics[width=0.65\textwidth]{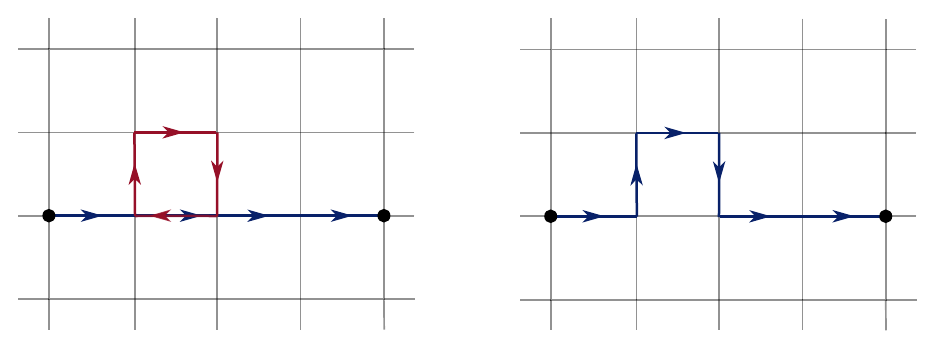}
    		\caption{Two separated static charges are joined, in the strong coupling limit, by an electric flux line. If the coupling constant decreases, pure-gauge loops will appear. If some link in the loop coincides with the electric flux line between the charges (in the opposite direction), it will result in a deformation of the flux line.}
    		\label{fig:static_charges}
\end{figure}

Consider now a gauge-invariant sector with non-zero static charges in some vertices (Fig. \ref{fig:static_charges}). In this case, the empty state is not the ground state of the system even in the strong coupling limit. The reason for this is that Gauss law (\ref{eq:gauss_law}) requires an electric flux line to exist between every pair of static charges, being the ground state in the strong coupling limit characterized by the shortest of those paths. If the value of the coupling constant $g$ decreases, closed loops will be formed all around the lattice, deforming the flux line between the static charges if they share a common link with the latter (Fig. \ref{fig:static_charges}).

\begin{figure}[t]
  			\centering
    		\includegraphics[width=0.65\textwidth]{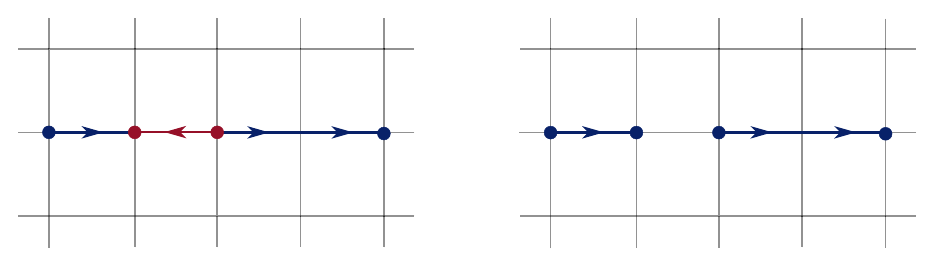}
    		\caption{If a meson-like excitation is created along the electric flux line of an existing mesonic state, the latter can break, giving birth to two separated and smaller ``mesons".}
    		\label{fig:string_breaking}
\end{figure}

If matter is present in the theory, the flux lines will not only fluctuate but can even break. For $R\neq 0$, the gauge-matter interaction term can create charge excitations in two nearest-neighbor vertices. We can find, then, states characterized by separated pairs of dynamical charges joined via fluctuating flux lines, which we will call \textit{mesons}. The difference with respect to the pure-gauge case is that, now, the flux lines can break, resulting in two new\textemdash and shorter\textemdash meson-like states (Fig. \ref{fig:string_breaking}). This phenomenon resembles the behaviour of QCD, where, due to the confinement mechanism, the flux lines between a quark-antiquark pair can break through the creation of a new pair from the vacuum (string breaking) \cite{Schwartz}.

Here, we have described some basic features of this lattice model. The complete phase diagram of the theory is described in \ref{app:action} for different dimensions and different representations of the symmetry group.

\section{Quantum Simulation}
\label{sec:quantum_simulation}

We shall now describe a quantum simulation scheme for the lattice Abelian-Higgs theory using ultracold bosonic atoms trapped in optical lattices. The goal is to impose certain conditions on this highly controllable atomic system, such that the Hamiltonian that describes it maps, approximately, the Hamiltonian under study (\ref{eq:my_Hamiltonian}). Both the parameters appearing in the latter, as well as the degree of approximation, can be controlled experimentally. This allows to study, for instance, the ground state of this many-body quantum theory in the interacting regime, which is a hard task to perform using standard analytical or numerical calculations.

The simulation proposal involves three main steps. First, starting from the most general multi-species Hamiltonian that describes ultracold bosons, the experimental requirements that are necessary for the simulation will be presented. After that, the transformations required for mapping the degrees of freedom of the atomic system to those of the Abelian-Higgs theory will be introduced. Finally, we will see how the desired Hamiltonian is effectively obtained, apart from correction terms, after increasing one of the energy scales of the system compared to the rest. Most of these corrections can be made as small as required by changing the experimental parameters of the atomic system, keeping the freedom to move through the phase space of the simulated theory, and by increasing the number of atoms used in the simulation.

\subsection{Ultracold Atoms in Optical Lattices}

\subsubsection{Atomic Hamiltonian}

Ultracold atomic gases consist of neutral atoms trapped and cooled down to almost absolute zero temperature, where quantum effects play a significant role. They represent one of the most relevant platforms for the study of collective quantum phenomena in regimes that are either hard to access experimentally, or where numerical simulations do not provide sufficiently good results. In particular, the so-called optical lattices\textemdash structures made of counter-propagating lasers\textemdash can be prepared in a way that emulates a periodic lattice structure with atoms bound to the vertices \cite{Jaksch,Lewenstein,Bloch}. Within these lattices, interactions among the atoms can be tuned in order to recreate Hamiltonian models describing many-body systems, both in condensed matter and in high energy physics. In this work, we will consider the most general multi-species bosonic Hamiltonian \cite{Lewenstein},
\begin{equation}
\label{eq:multi_Hubbard_model}
\hat{H}=\sum_{\alpha,\beta}\sum_{i,j} t_{i,j}^{\alpha,\beta}\hat{b}_{i,\alpha}^\dagger\hat{b}^{\vphantom{\dagger}}_{j,\beta}+\sum_{i,j,k,l}\sum_{\alpha,\beta,\delta,\gamma}U_{i,j,k,l}^{\alpha,\beta,\delta,\gamma}\hat{b}_{i,\alpha}^\dagger\hat{b}_{j,\beta}^\dagger\hat{b}^{\vphantom{\dagger}}_{k,\delta}\hat{b}^{\vphantom{\dagger}}_{l,\gamma}\,,
\end{equation}
where the latin indices denote the different sites of an optical lattice, and the greek indices represent different bosonic species (either different kinds of atoms or atoms with different internal levels). The operators $\hat{b}^{\vphantom{\dagger}}_{i,\alpha}$ and $\hat{b}^\dagger_{i,\alpha}$ are bosonic creation and annihilation operators, respectively, fulfilling the commutation relations
\begin{equation}
[\hat{b}^{\vphantom{\dagger}}_{i,\alpha},\hat{b}_{j,\beta}^\dagger]=\delta^{\vphantom{\dagger}}_{ij}\delta^{\vphantom{\dagger}}_{\alpha,\beta}\,.
\end{equation}

The parameters $t^{\alpha,\beta}_{i,j}$ and $U_{i,j,k,l}^{\alpha,\beta,\delta,\gamma}$ are constrained by conservation laws and their respective symmetries, and can be fine-tuned using experimental techniques such as (optical) Feshbach resonances \cite{Bloch,Chin,Fedichev,Theis}. The first term in (\ref{eq:multi_Hubbard_model}) corresponds to a hopping process of individual atoms along the lattice sites, while the second is a two-body collision process. For the latter, the conservation of the total angular momentum $\mathbf{F}$ implies, in particular, that the following condition must be fulfilled in each collision,
\begin{equation}
m_F(\alpha)+m_F(\beta)=m_F(\delta)+m_F(\gamma)\,,
\end{equation}
where $m_F(\alpha)$ represents the hyperfine angular momentum state for the atomic species $\alpha$.

\subsubsection{Experimental Requirements}

\begin{figure}[t]
  			\centering
    		\includegraphics[width=0.8\textwidth]{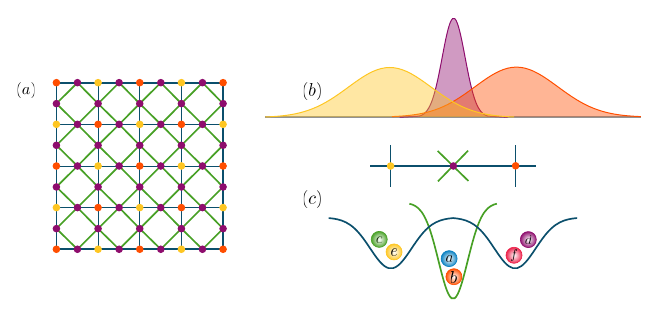}
    		\caption{(a) Two superimposed optical lattices, oriented with a $45\degree$ angle between their links, are required to simulate both the matter and gauge degrees of freedom. The vertices of one lattice (blue in the figure) represent the vertices of the simulated one, whereas the vertices of the rotated lattice (green) correspond the the links of the latter. The ``matter" lattice is filled with four types of bosons (c), two species per site, alternating on even and odd vertices. Two of these bosonic species represent dynamical matter (species $c$ and $d$), and the other two are auxiliary bosons required for the simulation (species $e$ and $f$). The ``gauge" lattice, on the other hand, is filled with two types of bosons on each site (species $a$ and $b$), with no distinction for even and odd links. The characteristics of these lattices must be configured such that there is no interactions between bosons from different sites of the ``gauge" lattice\textemdash using, for example, deep potentials\textemdash but allowing for interactions between them and the bosons from the nearest-neighbor vertices of the ``matter" lattice. This is the case if the corresponding Wannier functions have a non-vanishing intersection (b).}
    		\label{fig:super_figure}
\end{figure}

Our simulation proposal focuses on the $2+1$ dimensional case, which is the first ``non-trivial" dimension, since it contains plaquette interactions. In particular, we need two superimposed optical lattices to perform a lattice gauge theory simulation, whose minima correspond to the vertices and links of the simulated lattice, respectively (Fig. \ref{fig:super_figure}a). Filling and combining two such lattices is possible in the two-dimensional case, however, it becomes more complicated in three dimensions. All the theoretical calculations are valid, nevertheless, for $d+1$ dimensions, with $d\geq 2$.

To simulate the complete Abelian-Higgs Hamiltonian (\ref{eq:my_Hamiltonian}) we need six different bosonic species (Fig. \ref{fig:super_figure}c). Two of them, $a$ and $b$, are placed on the links of the simulated lattice and represent the gauge degrees of freedom. Another pair of bosons simulates the scalar fields on the vertices (dynamical matter), one species on the odd vertices, $c$, and the other one on the even ones, $d$. The last two species, $e$ and $f$, are distributed, again, on odd and even vertices, respectively, acting as auxiliary particles required to obtain the plaquette interactions in an effective way, as we will see. Each pair of bosonic species corresponds to the same type of atom, but their internal level differ in the value of the hyperfine angular momentum, $m_F$. We will denote both types of dynamical bosons with $\eta$, and both auxiliary ones with $\chi$, in those situations where a distinction between the hyperfine angular momentum level is not important.

\vspace{0.5cm}

\begin{center}
\begin{tabular}{l*{6}{c}r}
	           &  &\,\, Dyn. $\eta$\,\,\,  & Aux. $\chi$ \\
\hline
& Odd & $c$ & $e$  \\
& Even & $d$ & $f$  \\
\end{tabular}
\end{center}

\subsection{Primitive Hamiltonian}

For dimensions $d+1$, with $d>1$, the Hamiltonians that describe lattice gauge theories, such as the Abelian-Higgs one (\ref{eq:my_Hamiltonian}), include plaquette-type interactions made out of the product of four unitary operators (Fig. \ref{fig:plaquette}). Such four-body interactions are not found in the ultracold atomic Hamiltonian. In order to simulate them, we have to obtain these type of terms effectively \cite{Erez_review,Erez_5}, as we will explain in later sections.

We shall call \textit{primitive} Hamiltonian the one from which the desired Abelian-Higgs Hamiltonian can be obtained effectively, acting only on a low-energy sector. It has the form
\begin{equation}
\label{eq:primitive_hamiltonian_unitary}
\begin{aligned}
&\hat{H}=\lambda \sum_{\mathbf{n}} \hat{N}^\chi_{\mathbf{n}}(\hat{N}^\chi_{\mathbf{n}}-1)+\epsilon\sum_{\mathbf{n},k}\left[\hat{\chi}^\dagger_{\mathbf{n}\vphantom{\hat{k}}}\hat{U}^\dagger_{\mathbf{n}\vphantom{\hat{k}},k}\hat{\chi}^{\vphantom{\dagger}}_{\mathbf{n}+\hat{k}}+h.c.\right]\\&+\mu \sum_{\mathbf{n},k} \hat{E}^2_{\mathbf{n},k}+\mu^\prime \sum_{\mathbf{n}} \hat{Q}^2_{\mathbf{n}}+\epsilon^\prime\sum_{\mathbf{n},k}\left[\hat{\phi}^\dagger_{\mathbf{n}\vphantom{\hat{k}}}\hat{U}^\dagger_{\mathbf{n}\vphantom{\hat{k}},k}\hat{\phi}^{\vphantom{\dagger}}_{\mathbf{n}+\hat{k}}+h.c.\right].
\end{aligned}
\end{equation}
The first row contains the operators corresponding to auxiliary bosons living on the vertices of the lattice. $\hat{\chi}^\dagger_{\mathbf{n}}$ and $\hat{\chi}^{\vphantom{\dagger}}_{\mathbf{n}}$ are the bosonic creation and annihilation operators, respectively, and $\hat{N}^\chi_{\mathbf{n}}=\hat{\chi}^\dagger_{\mathbf{n}}\hat{\chi}^{\vphantom{\dagger}}_{\mathbf{n}}$ is the corresponding number operator. The second row contains all the terms of (\ref{eq:my_Hamiltonian}) except for the plaquette interaction. The primitive Hamiltonian is invariant as well under $U(1)$ local transformations applied on the vertices of the lattice.

We will now focus on bringing the atomic Hamiltonian to the primitive (intermediate) form (\ref{eq:primitive_hamiltonian_unitary}). Although only an approximate Hamiltonian can be obtained, the level of approximation can be controlled by using a different number of atoms in the simulation.

\subsection{Operator Transformations}

The first step towards a quantum simulation is to construct a one-to-one mapping between the degrees of freedom of both the simulating and the simulated systems. Here, this is achieved by defining the operators that appear in the Abelian-Higgs Hamiltonian (\ref{eq:my_Hamiltonian}) in terms of bosonic creation and annihilation operators corresponding to the different atomic species. For a finite number of atoms trapped on the vertices and links of the optical lattice, the correspondence is not exact, in the sense that the defined operators will act only on a truncated  Hilbert space. However, the approximation will improve when the number of atoms increases, as the connection between both systems becomes exact in the infinite number of atoms limit.

On the vertices of the lattice (both even and odds), we introduce the operators $\hat{Q}_{\mathbf{n}}$, $\hat{\phi}_{\mathbf{n}}$ and $\hat{\phi}^\dagger_{\mathbf{n}}$ using the second-quantized operators $\hat{\eta}_{\mathbf{n}}$ and $\hat{\eta}^\dagger_{\mathbf{n}}$ for the dynamical bosons,
\begin{equation}
\label{eq:transformation_c}
\begin{aligned}
\hat{\eta}_{\mathbf{n}}&=\hat{\phi}_{\mathbf{n}}\left(N_{0,v}+\hat{Q}_{\mathbf{n}}\right)^{1/2}, &
\hat{\eta}^\dagger_{\mathbf{n}}&=\left(N_{0,v}+\hat{Q}_{\mathbf{n}}\right)^{1/2}\hat{\phi}^\dagger_{\mathbf{n}}
\end{aligned}
\end{equation}
and
\begin{equation}
\label{eq:charge_operators}
\hat{Q}_{\mathbf{n}}=\hat{\eta}^\dagger_{\mathbf{n}}\hat{\eta}^{\vphantom{\dagger}}_{\mathbf{n}}-N_{0,v}\,,
\end{equation}
where $N_{0,v}$ is the density of $\eta$ bosons. With these relations, both the canonical commutation relations  for the bosonic operators and the commutation relations (\ref{eq:commutation_phi_Q}) are fulfilled. According to this definition, the operator $\hat{Q}_{\mathbf{n}}$ is bounded from below by $-N_{0,v}$. For this reason, it is not exactly equivalent to the dynamical charge operator that appears in (\ref{eq:my_Hamiltonian}), however, we can think about it as the latter acting on a truncated Hilbert space\textemdash the one spanned by the basis $\ket{m}$ with
$m\in\mathbb{Z}\cap[-N_{0,v},\infty)$\textemdash where it is equivalent to the dynamical charge operator. For this reason, we use the same notation for both, understanding, in the following, that we are referring to the operator acting on the truncated space.

Another problem concerning the finiteness of $N_{0,l}$ is that, by imposing the commutation relations (\ref{eq:commutation_phi_Q}), $\hat{\eta}^{\vphantom{\dagger}}_{\mathbf{n}}$ and $\hat{\eta}^\dagger_{\mathbf{n}}$ cannot be unitary, which also means that they are not the exponential of a self-adjoint phase $\hat{\varphi}_{\mathbf{n}}$,  $\hat{\phi}_{\mathbf{n}}=e^{-i\hat{\varphi}_{\mathbf{n}}}$. This is related to the \textit{quantum phase operator} problem  \cite{Phase_0,Phase_1,Phase_2,Phase_3,Phase_4}, and lies in the fact that $\hat{Q}_{\mathbf{n}}$ is bounded from below. A unitary operator is well-defined only in the limit $N_{0,v}\rightarrow \infty$, such that the charge operator is no longer bounded.

We are working with a large, but finite, number of atoms $N_{0,v}$ on each vertex. The reason it is still fine, from a physical point of view, to use these definitions\textemdash although they are not formally correct\textemdash is that in all the interaction terms of the Hamiltonian the bosonic operators appear in pairs, corresponding to nearest-neighbor vertices, $\hat{\eta}^\dagger_{\mathbf{n}\vphantom{+\hat{k}}}\hat{\eta}^{\vphantom{\dagger}}_{\mathbf{n}+\hat{k}}$. Therefore, only phase differences between different vertices are relevant. The phase difference is a well-defined quantity, since it is canonically conjugate to a number operator, $\hat{N}=\hat{N}^c-\hat{N}^d$, which is not bounded from below.

The operators associated to the gauge fields, $\hat{E}_{\mathbf{n},k}$ and $\hat{U}_{\mathbf{n},k}$ can be obtained from the bosonic operators on the links if we first represent them in a similar way as the atomic operators on the vertices,
\begin{equation}
\begin{aligned}
\hat{a}_{\mathbf{n},k}&=\hat{U}^a_{\mathbf{n},k}\left(\frac{N_{0,l}}{2}+\hat{E}^a_{\mathbf{n},k}\right)^{1/2}, & \hat{b}_{\mathbf{n},k}&=\hat{U}^b_{\mathbf{n},k}\left(\frac{N_{0,l}}{2}+\hat{E}^b_{\mathbf{n},k}\right)^{1/2},
\end{aligned}
\end{equation}
where $\hat{U}^a_{\mathbf{n},k}$ and $\hat{U}^b_{\mathbf{n},k}$ are lowering operators, and
\begin{equation}
\begin{aligned}
\hat{E}^a_{\mathbf{n},k}&=\hat{a}^\dagger_{\mathbf{n},k}\hat{a}^{\vphantom{\dagger}}_{\mathbf{n},k}-\frac{N_{0,l}}{2}\,, & \hat{E}^b_{\mathbf{n},k}&=\hat{b}^\dagger_{\mathbf{n},k}\hat{b}^{\vphantom{\dagger}}_{\mathbf{n},k}-\frac{N_{0,l}}{2}\,.
\end{aligned}
\end{equation}
Again, if the commutation relations (\ref{eq:commutator_U_E}) are fulfilled, the canonical relations for the bosonic operators will be satisfied.

The total number of bosons on each link, $N_{0,l}$ is a conserved quantity,
\begin{equation}
N_{0,l}=\hat{a}^\dagger_{\mathbf{n},k}\hat{a}^{\vphantom{\dagger}}_{\mathbf{n},k}+\hat{b}^\dagger_{\mathbf{n},k}\hat{b}^{\vphantom{\dagger}}_{\mathbf{n},k}\,,
\end{equation}
since, as we shall see, only products of bosonic operators of the form $\hat{a}^\dagger_{\mathbf{n},k}\hat{b}^{\vphantom{\dagger}}_{\mathbf{n},k}$ and $\hat{b}^\dagger_{\mathbf{n},k}\hat{a}^{\vphantom{\dagger}}_{\mathbf{n},k}$ will appear in the Hamiltonian. This implies that $\hat{E}^a_{\mathbf{n},k}=-\hat{E}^b_{\mathbf{n},k}\equiv \hat{E}_{\mathbf{n},k}$. We use latter as the electric field operator on the link $(\mathbf{n},k)$, which can be expressed also as
\begin{equation}
\label{eq:electric_field_difference}
E_{\mathbf{n},k}=\frac{1}{2}\left(\hat{a}^\dagger_{\mathbf{n},k}\hat{a}^{\vphantom{\dagger}}_{\mathbf{n},k}-\hat{b}^\dagger_{\mathbf{n},k}\hat{b}^{\vphantom{\dagger}}_{\mathbf{n},k}\right).
\end{equation}

We construct the unitary operators acting on the links as
\begin{equation}
\hat{U}^{\dagger}_{\mathbf{n},k}=\hat{U}^{a\,\dagger}_{\mathbf{n},k}\hat{U}^{b\vphantom{\dagger}}_{\mathbf{n},k}\,,
\end{equation}
satisfying the correct commutation relations (\ref{eq:commutator_U_E}) with the electric field (\ref{eq:electric_field_difference}). Note that, in this case, the operator $E_{\mathbf{n},k}$ is not bounded from below. $\hat{U}^{\vphantom{\dagger}}_{\mathbf{n},k}$ and $\hat{U}^\dagger_{\mathbf{n},k}$ are, therefore, well-defined unitary operators, although they only act on a truncated Hilbert space that grows with $N_{0,l}$.

Finally, the product of two bosonic operators on the links can be written as
\begin{equation}
\label{eq:transformation_ab}
\begin{aligned}
\hat{a}^\dagger_{\mathbf{n},k}\hat{b}^{\vphantom{\dagger}}_{\mathbf{n},k}&=\hat{U}^{\dagger}_{\mathbf{n},k}\left(\frac{N_{0,l}(N_{0,l}+2)}{4}-\hat{E}_{\mathbf{n},k}(\hat{E}_{\mathbf{n},k}+1)\right)^{1/2},\\
\hat{b}^\dagger_{\mathbf{n},k}\hat{a}^{\vphantom{\dagger}}_{\mathbf{n},k}&=\left(\frac{N_{0,l}(N_{0,l}+2)}{4}-\hat{E}_{\mathbf{n},k}(\hat{E}_{\mathbf{n},k}+1)\right)^{1/2}\hat{U}^{\vphantom{\dagger}}_{\mathbf{n},k}\,,
\end{aligned}
\end{equation}
where we used the property $\hat{U}\,f(\hat{E})=f(\hat{E}+1)\,\hat{U}$ (see \ref{app:non_commuting}). In the following, we will assume this property\textemdash similarly for the operators on the vertices, $\hat{\phi}\,f(\hat{Q})=f(\hat{Q}+1)\,\hat{\phi}$\textemdash to deal with these non-commuting operators.

In order to simplify the notation, we define the non-unitary operators
\begin{equation}
\label{eq:non_unitary}
\begin{aligned}
\hat{\Phi}^\dagger_{\mathbf{n}}&\equiv\left(1+\frac{1}{N_{0,v}}\hat{Q}_\mathbf{n}\right)^{1/2}\hat{\phi}^\dagger_{\mathbf{n}}\,,\\
\hat{\mathcal{U}}^\dagger_{\mathbf{n},k}&\equiv\left(1-\frac{4}{N_{0,l}(N_{0,l}+2)}(\hat{E}^2_{\mathbf{n},k}-\hat{E}_{\mathbf{n},k})\right)^{1/2}\hat{U}^{\dagger}_{\mathbf{n}\vphantom{\hat{k}},k}\,,
\end{aligned}
\end{equation}
which become the desired unitary ones, $\hat{\phi}^{\dagger}_{\mathbf{n}}$ and $\hat{U}^{\dagger}_{\mathbf{n}}$, in the limits $N_{0,v}\rightarrow\infty$ and $N_{0,l}\rightarrow\infty$, respectively. The connection with the atomic operators is summarized in the following relations,
\begin{equation}
\label{eq:atomic_gauge_connection}
\begin{aligned}
\hat{\Phi}^\dagger_{\mathbf{n}}&=\frac{\hat{\eta}^\dagger_{\mathbf{n}}}{\sqrt{N_{0,v}}}\,,
 & \hat{\mathcal{U}}^\dagger_{\mathbf{n},k}&=\frac{\hat{a}^\dagger_{\mathbf{n},k}\hat{b}^{\vphantom{\dagger}}_{\mathbf{n},k}}{\sqrt{N_{0,l}(N_{0,l}+2)/4}}\,,\\
\hat{\Phi}^{\vphantom{\dagger}}_{\mathbf{n}}&=\frac{\hat{\eta}^{\vphantom{\dagger}}_{\mathbf{n}}}{\sqrt{N_{0,v}}}\,, & \hat{\mathcal{U}}^{\vphantom{\dagger}}_{\mathbf{n},k}&=\frac{\hat{b}^\dagger_{\mathbf{n},k}\hat{a}^{\vphantom{\dagger}}_{\mathbf{n},k}}{\sqrt{N_{0,l}(N_{0,l}+2)/4}}\,.
\end{aligned}
\end{equation}

\subsection{Gauge-Invariant Interactions}

Local gauge invariance is not a fundamental symmetry in a system of ultracold atoms. It is essential, however, for the description of the interactions present in high energy physics, as we have seen. If one aims to correctly describe the latter, gauge symmetry must be imposed on the simulating system. This can be done either in a digital \cite{digital_gauge_1} or in an analog simulation framework. For the latter, there are two distinct ways to achieve this \cite{Erez_review}, one where gauge invariance is obtained effectively, as an emerging, low energy symmetry \cite{Erez_1,Erez_2,Erez_3}, and another where it is mapped, exactly, to a fundamental symmetry of the system\textemdash such as the conservation of hyperfine angular momentum in atomic collisions. In this case, as opposed to an effective gauge symmetry, gauge invariance is exact, making the simulation more robust against experimental imperfections. Using the second approach, quantum simulations of both abelian and non-abelian lattice gauge theories have been proposed \cite{Erez_4,Erez_5}. Here, we will employ it to simulate the $U(1)$ gauge invariance of the Abelian-Higgs model.

\begin{figure}[t]
  			\centering
    		\includegraphics[width=0.45\textwidth]{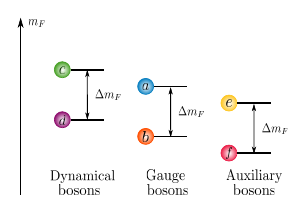}
    		\caption{The hyperfine angular momenta of the six bosonic species must fulfilled certain conditions, such that the conservation of this quantity during atomic collisions permits the desire atomic interactions.}
    		\label{fig:hyperfine_momentum}
\end{figure}

\begin{figure}[t]
  			\centering
    		\includegraphics[width=0.65\textwidth]{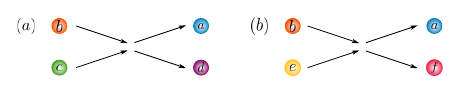}
    		\caption{The conservation of the total hyperfine angular momentum in the atomic collisions generates the desired gauge-invariant interactions between the dynamical (a) and auxiliary matter (b) on the vertices of the lattice, and the gauge degrees of freedom on the links.}
    		\label{fig:collision_2}
		\end{figure}

In order to obtain the correct ``gauge-invariant" interactions, the hyperfine angular momenta of the six bosonic species must fulfill certain conditions (Fig. \ref{fig:hyperfine_momentum}),
\begin{equation}
m_{F}(a)-m_{F}(b)=m_{F}(c)-m_{F}(d)=m_{F}(e)-m_{F}(f)\equiv \Delta m\,.
\end{equation}
Since only collision processes that conserve the total angular momentum are allowed, this condition selects the type of terms that appear in the atomic Hamiltonian. Consider, for instance, the interactions between a dynamical boson on a vertex and a gauge boson on one of its nearest-neighbor links. The conservation of hyperfine angular momentum allows both species-changing collisions (Fig. \ref{fig:collision_2}a),
\begin{equation}
\label{eq:species_changing}
\hat{d}^{\dagger}_{\mathbf{n}\vphantom{\hat{k}}}\hat{a}^\dagger_{\mathbf{n}\vphantom{\hat{k}},k}\hat{b}^{\vphantom{\dagger}}_{\mathbf{n}\vphantom{\hat{k}},k}\hat{c}^{\vphantom{\dagger}}_{\mathbf{n}+\hat{k}}+\hat{c}^{\dagger}_{\mathbf{n}+\hat{k}}\hat{b}^\dagger_{\mathbf{n}+\hat{k},k}\hat{a}^{\vphantom{\dagger}}_{\mathbf{n}+\hat{k},k}\hat{d}^{\vphantom{\dagger}}_{\mathbf{n}+2\hat{k}}\,,
\end{equation}
and non-species-changing ones,
\begin{equation}
\begin{aligned}
&\hat{d}^{\dagger}_{\mathbf{n}\vphantom{\hat{k}}}\hat{d}^{\vphantom{\dagger}}_{\mathbf{n}\vphantom{\hat{k}}}\left(\hat{a}^{\dagger}_{\mathbf{n},k}\hat{a}^{\vphantom{\dagger}}_{\mathbf{n},k}+\hat{b}^{\dagger}_{\mathbf{n},k}\hat{b}^{\vphantom{\dagger}}_{\mathbf{n},k}\right)=\hat{d}^{\dagger}_{\mathbf{n}\vphantom{\hat{k}}}\hat{d}^{\vphantom{\dagger}}_{\mathbf{n}\vphantom{\hat{k}}}N_{0,l}\\
&\hat{c}^{\dagger}_{\mathbf{n}+\hat{k}}\hat{c}^{\vphantom{\dagger}}_{\mathbf{n}+\hat{k}}\left(\hat{a}^{\dagger}_{\mathbf{n},k}\hat{a}^{\vphantom{\dagger}}_{\mathbf{n},k}+\hat{b}^{\dagger}_{\mathbf{n},k}\hat{b}^{\vphantom{\dagger}}_{\mathbf{n},k}\right)=\hat{c}^{\dagger}_{\mathbf{n}+\hat{k}}\hat{c}^{\vphantom{\dagger}}_{\mathbf{n}+\hat{k}}N_{0,l}\,.
\end{aligned}
\end{equation}
The sum of these terms becomes a constant, proportional to the number of bosons, when summed over all vertices, if the scattering lengths are properly tuned. The same kind of processes appears for the interaction between gauge and auxiliary bosons (Fig. \ref{fig:collision_2}b).

The species-changing collisions can be seen to be equivalent to the matter-gauge interactions in (\ref{eq:my_Hamiltonian}), after applying the canonical transformations
\begin{equation}
\label{eq:canonical_transformation}
\left(
\begin{array}{c}
\hat{a}_{\mathbf{n}+\hat{k}}\\
\hat{b}_{\mathbf{n}+\hat{k}}
\end{array}
\right)\rightarrow\hat{\sigma}_x^{n_1+n_2}
\left(
\begin{array}{c}
\hat{a}_{\mathbf{n}+\hat{k}}\\
\hat{b}_{\mathbf{n}+\hat{k}}
\end{array}
\right)\,.
\end{equation}
Both $c$ and $d$ operators can be denoted now by $\eta$, and both $e$ and $f$ by $\chi$. Using the transformations defined in the last section (\ref{eq:atomic_gauge_connection}), the species-changing collisions for the dynamical bosons are just
\begin{equation}
\begin{aligned}
\epsilon^\prime_0 \left(\hat{\eta}^\dagger_{\mathbf{n}\vphantom{\hat{k}}}\hat{a}^\dagger_{\mathbf{n}\vphantom{\hat{k}},k}\hat{b}^{\vphantom{\dagger}}_{\mathbf{n}\vphantom{\hat{k}},k}\hat{\eta}^{\vphantom{\dagger}}_{\mathbf{n}+\hat{k}}+h.c.\right)=\,&\epsilon^\prime \left(\hat{\Phi}^\dagger_{\mathbf{n}\vphantom{\hat{k}}}\hat{\mathcal{U}}^{\dagger}_{\mathbf{n}\vphantom{\hat{k}},k}\hat{\Phi}^{\vphantom{\dagger}}_{\mathbf{n}+\hat{k}}+h.c.\right),
\end{aligned}
\end{equation}
where a new coupling constant has been defined,
\begin{equation}
\epsilon^\prime=\epsilon^\prime_0 N^{\vphantom{\prime}}_{0,v}\sqrt{\frac{N_{0,l}(N_{0,l}+2)}{4}}\,.
\end{equation}

For the auxiliary bosons, the species-changing collisions transform to
\begin{equation}
\epsilon_0\left(\hat{\chi}^\dagger_{\mathbf{n}\vphantom{\hat{k}}}\hat{a}^\dagger_{\mathbf{n}\vphantom{\hat{k}},k}\hat{b}^{\vphantom{\dagger}}_{\mathbf{n}\vphantom{\hat{k}},k}\hat{\chi}^{\vphantom{\dagger}}_{\mathbf{n}+\hat{k}}+h.c.\right)=\,\epsilon\left(\hat{\chi}^\dagger_{\mathbf{n}\vphantom{\hat{k}}}\hat{\mathcal{U}}^{\dagger}_{\mathbf{n}\vphantom{\hat{k}},k}\hat{\chi}^{\vphantom{\dagger}}_{\mathbf{n}+\hat{k}}+h.c.\right),
\end{equation}
with
\begin{equation}
\epsilon=\epsilon_0\sqrt{\frac{N_{0,l}(N_{0,l}+2)}{4}}\,.
\end{equation}

Note that these kind of interactions are possible provided the Wannier functions of the corresponding bosons on the lattice overlap (Fig. \ref{fig:super_figure}b).

The conservation of hyperfine angular momentum allows the two types of bosons to interact as well. This would result in non-desired terms in the simulated Hamiltonian. For this reason, the system must be prepared such that this kind of interactions does not occur, using, for example, Feshbach resonances to reduce the corresponding scattering length.

Consider now the non-interacting terms in (\ref{eq:primitive_hamiltonian_unitary}). Both the charge and the electric field kinetic terms arise from on-site collisions. For the vertices we have
\begin{equation}
\mu^\prime\sum_{\mathbf{n}}\hat{\eta}^\dagger_{\mathbf{n}}\hat{\eta}^{\vphantom{\dagger}}_{\mathbf{n}}\hat{\eta}^\dagger_{\mathbf{n}}\hat{\eta}^{\vphantom{\dagger}}_{\mathbf{n}}=\mu^\prime\sum_{\mathbf{n}}\hat{Q}^2_{\mathbf{n}}+\mathrm{``constants"},
\end{equation}
where we used the conservation of the total number of bosons on the vertices.

On the links, different on-site collision processes appear. They correspond to the electric field kinetic term if the right couplings are chosen,
\begin{equation}
\begin{aligned}
&\frac{\mu}{4}\left(\hat{a}^\dagger_{\mathbf{n},k}\hat{a}^{\vphantom{\dagger}}_{\mathbf{n},k}\hat{a}^\dagger_{\mathbf{n},k}\hat{a}^{\vphantom{\dagger}}_{\mathbf{n},k}+\hat{b}^\dagger_{\mathbf{n},k}\hat{b}^{\vphantom{\dagger}}_{\mathbf{n},k}\hat{b}^\dagger_{\mathbf{n},k}\hat{b}^{\vphantom{\dagger}}_{\mathbf{n},k}-2\hat{a}^\dagger_{\mathbf{n},k}\hat{a}^{\vphantom{\dagger}}_{\mathbf{n},k}\hat{b}^\dagger_{\mathbf{n},k}\hat{b}^{\vphantom{\dagger}}_{\mathbf{n},k}\right)\\
&=\frac{\mu}{4}\left(\hat{a}^\dagger_{\mathbf{n},k}\hat{a}^{\vphantom{\dagger}}_{\mathbf{n},k}-\hat{b}^\dagger_{\mathbf{n},k}\hat{b}^{\vphantom{\dagger}}_{\mathbf{n},k}\right)^2=\mu\hat{E}^2_{\mathbf{n},k}\,.
\end{aligned}
\end{equation}

Finally, the hard-core constraint for the auxiliary bosons can be obtained from on-site collision among them, as well as atomic mass terms,
\begin{equation}
\lambda \sum_{\mathbf{n}} \left(\hat{\chi}^\dagger_{\mathbf{n}}\hat{\chi}^{\vphantom{\dagger}}_{\mathbf{n}} \hat{\chi}^\dagger_{\mathbf{n}}\hat{\chi}^{\vphantom{\dagger}}_{\mathbf{n}}-\hat{\chi}^\dagger_{\mathbf{n}}\hat{\chi}^{\vphantom{\dagger}}_{\mathbf{n}}\right)=\lambda \sum_{\mathbf{n}}\sum_{\mathbf{n}} \hat{N}^\chi_{\mathbf{n}}(\hat{N}^\chi_{\mathbf{n}}-1)\,.
\end{equation}

Collecting all the building blocks we see how the atomic Hamiltonian corresponds to
\begin{equation}
\label{eq:primitive_hamiltonian_app1}
\begin{aligned}
&H=\lambda \sum_{\mathbf{n}} \hat{N}^\chi_{\mathbf{n}}(\hat{N}^\chi_{\mathbf{n}}-1)+\epsilon \sum_{\mathbf{n},k} \left[\hat{\chi}^\dagger_{\mathbf{n}\vphantom{\hat{k}}}\hat{\mathcal{U}}^{\dagger}_{\mathbf{n}\vphantom{\hat{k}},k}\hat{\chi}^{\vphantom{\dagger}}_{\mathbf{n}+\hat{k}}+h.c.\right]\\
&+\mu \sum_{\mathbf{n},k} \hat{E}^2_{\mathbf{n},k}+\mu^\prime \sum_{\mathbf{n}} \hat{Q}^2_{\mathbf{n}}+\epsilon^\prime \sum_{\mathbf{n},k} \left[\hat{\Phi}^\dagger_{\mathbf{n}\vphantom{\hat{k}}}\hat{\mathcal{U}}^{\dagger}_{\mathbf{n}\vphantom{\hat{k}},k}\hat{\Phi}^{\vphantom{\dagger}}_{\mathbf{n}+\hat{k}}+h.c.\right],
\end{aligned}
\end{equation}
which, in the limit $N_{0,l},N_{0,v}\rightarrow\infty$ results in (\ref{eq:primitive_hamiltonian_unitary}). In an experiment, only a finite number of atoms is placed on each link and vertex. However, this Hamiltonian provides a good approximation, even for small number of atoms \cite{Kasper}. We will discuss the effect that the number of atoms have on the simulation after obtaining the effective Hamiltonian.

\subsection{Effective Hamiltonian}
\label{sec:effective}

\subsubsection{Hard-Core Bosons}

We shall next describe how the plaquette interactions can be obtained in an effective way from the primitive Hamiltonian (\ref{eq:primitive_hamiltonian_app1}). The main idea is to impose a strong energy penalty in the atomic Hamiltonian, in such a way that the Abelian-Higgs Hamiltonian (\ref{eq:my_Hamiltonian}) is obtained as a low energy effective approximation, including the plaquette interactions \cite{Erez_5}. The advantage of this method is that the building blocks from which the plaquettes are constructed are already gauge-invariant. Therefore, even if the effective description contains undesired terms, these will not violate the gauge symmetry.

The method requires the use of auxiliary particles, which can be either fermions or bosons \cite{Erez_review}. In this case, we make use of \textit{hardcore} auxiliary bosons sitting on the vertices of the lattice, which we have already introduced in the primitive Hamiltonian (\ref{eq:primitive_hamiltonian_app1}). The energy penalty comes from the on-site interaction between auxiliary bosons,
\begin{equation}
\label{eq:penalty_term}
H_{\mathrm{HC}}=\lambda \sum_{\mathbf{n}} \hat{N}^\chi_{\mathbf{n}}(\hat{N}^\chi_{\mathbf{n}}-1)\,.
\end{equation}

\begin{figure}[t]
  			\centering
    		\includegraphics[width=0.25\textwidth]{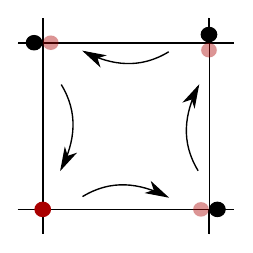}
    		\caption{The plaquette interactions are obtained as fourth order correction in the effective Hamiltonian. A virtual process consisting on one auxiliary particle hopping four times along the links of a plaquette, and returning to its original positions, results in the correct product of four unitary operators, each one attached to one link, that gives rise to the plaquette term.}
    		\label{fig:hopping_plaquette}
\end{figure}

We initialize the system with one auxiliary boson per vertex. If $\lambda$ is much larger than $\epsilon$, the hopping of auxiliary bosons to neighboring vertices implies a large energy penalty. The effective Hamiltonian acting on the sector characterized by one boson per vertex contains all the terms in the primitive Hamiltonian (\ref{eq:primitive_hamiltonian_app1}) that commutes with the penalty (\ref{eq:penalty_term}). It also includes higher order corrections, related to the virtual hopping of auxiliary bosons to neighboring vertices, and the return to their original position. In the fourth order, an auxiliary boson can hop along the links of a plaquette (Fig. \ref{fig:hopping_plaquette}), resulting the following effective contribution (\ref{app:calculations}),
\begin{equation}
\label{eq:effective_plaquette}
\begin{aligned}
&-\frac{5}{2}\frac{\epsilon^4}{\lambda^3}\left(\hat{\mathcal{U}}^{\vphantom{\dagger}}_{\mathbf{n}\vphantom{+\hat{k}},i}\hat{\mathcal{U}}^{\vphantom{\dagger}}_{\mathbf{n}+\hat{i},k}\hat{\mathcal{U}}^\dagger_{\mathbf{n}+\hat{k},i}\hat{\mathcal{U}}^\dagger_{\mathbf{n}\vphantom{+\hat{k}},k}+h.c.\right), & i\neq k,
\end{aligned}
\end{equation}
which is the desired plaquette interaction. Terms of the type $\epsilon^{n}/\lambda^{n-1}$ only contribute for even values of $n$. Therefore, the latter is, in practice, a second order correction. Apart from the plaquette term, many other terms appear in the effective expansion. Most of them arise as an effect of the non-unitary behaviour of the operators (\ref{eq:non_unitary}) when the number of atoms is finite, vanishing in the large number of atoms limit.

The use of hard-core bosons has several advantages, compared to obtaining the effective plaquettes using auxiliary fermions \cite{Erez_5}. First, the penalty term should be easier to implement experimentally, since it is homegeneous across the vertices of the lattice. Also, as we will see when we apply it to the complete lattice Abelian-Higgs theory, no undesired divergence terms or inhomogeneous renormalizations appear in the effective Hamiltonian, as opposed to in the fermionic case. Finally, the quantum simulation will only require bosonic atoms, rather than bosons and fermions, which simplifies as well the experiment.

\subsubsection{Effective Expansion}

Let us derive the complete effective Hamiltonian that acts on the sector of the energy constraint characterized by just one auxiliary boson per vertex. We assume that $\lambda$ is large compared to the rest of the coupling parameters,
\begin{equation}
\lambda\gg \epsilon,\,\epsilon^\prime,\,\mu,\,\mu^\prime.
\end{equation}
This can be achieved experimentally using Feshbach resonances. In order to calculate the effective Hamiltonian, we use time-independent perturbation theory, following \cite{eff_ham_1,eff_ham_2}. The effective Hamiltonian is perturbatively expanded in terms of $1/\lambda$. More details about the calculations can be found in \ref{app:calculations}. Here, we present the results up to fourth order. Notice that all the terms obtained in the effective expansion are gauge invariant, since the symmetry was exact in the primitive Hamiltonian (\ref{eq:primitive_hamiltonian_app1}).

The first order contributions to the effective Hamiltonian include all the terms in the primitive one (\ref{eq:primitive_hamiltonian_app1}) that commute with the penalty term (\ref{eq:penalty_term}),
\begin{equation}
\label{eq:H_eff_1.0}
\begin{aligned}
\hat{H}^{(1)}_\mathrm{eff}&=\mu \sum_{\mathbf{n},k} \hat{E}^2_{\mathbf{n},k}+\mu^\prime \sum_{\mathbf{n}} \hat{Q}^2_{\mathbf{n}}+\epsilon^\prime \sum_{\mathbf{n},k} \left(\hat{\Phi}^\dagger_{\mathbf{n}\vphantom{\hat{k}}}\hat{\mathcal{U}}^{\dagger}_{\mathbf{n}\vphantom{\hat{k}},k}\hat{\Phi}^{\vphantom{\dagger}}_{\mathbf{n}+\hat{k}}+h.c.\right).
\end{aligned}
\end{equation}

The second order contribution is a renormalization of the electric part of the Hamiltonian,
\begin{equation}
\hat{H}^{(2)}_\mathrm{eff}=\frac{\epsilon^2}{\lambda}\frac{4}{N_{0,l}(N_{0,l}+2)}\sum_{\mathbf{n},k} \hat{E}^2_{\mathbf{n},k}\,.
\end{equation}

In the third and fourth order, we get, apart from different renormalizations of the first order terms (\ref{eq:H_eff_1.0}), new contributions that were not present in the primitive Hamiltonian (\ref{eq:primitive_hamiltonian_app1}). The most important one is the already mentioned plaquette interaction (\ref{eq:effective_plaquette}). Due to the non-unitary character of the operators (\ref{eq:non_unitary}), the effective expansion provides, as well, a plethora of interaction terms that do not appear in the simulated Hamiltonian (\ref{eq:my_Hamiltonian}). Most of these corrections can be made negligible (see \ref{app:calculations}). The only extra terms that could affect the simulation in some regimes are the following fourth order contributions,
\begin{equation}
\begin{aligned}
&\hat{H}_\mathrm{ex}\equiv\frac{\epsilon^4}{\lambda^3}\frac{4}{N_{0,l}(N_{0,l}+2)}\left(\frac{\mu^2}{\epsilon^2}+\frac{2}{N_{0,l}(N_{0,l}+2)}\right)\sum_{\mathbf{n}} \hat{E}^4_{\mathbf{n},k}\\
&-\frac{\epsilon^4}{\lambda^3}\frac{2}{3}\frac{1}{\left(N_{0,l}(N_{0,l}+2)\right)^2}\sum_{\mathrm{``n.n."}}\left[\hat{E}_{\mathbf{n},k}\hat{E}_{\mathbf{n}^\prime,k^\prime}-11\hat{E}^2_{\mathbf{n},k}\hat{E}^2_{\mathbf{n}^\prime,k^\prime}\right],
\end{aligned}
\end{equation}
where the last term involves an interaction between nearest-neighbor links. The importance of the correction terms is analized now for different situations.

\subsection{Pure-Gauge Theory}

Before discussing the complete effective Hamiltonian, let us consider the pure-gauge theory. Most of the discussion generalizes when dynamical matter is present. Here, the situation corresponds to a quantum simulation of the Kogut-Susskind Hamiltonian \cite{Kogut-Susskind},
\begin{equation}
\label{eq:kogut_susskind_unitaries}
\hat{H}=\frac{g^2}{2}\sum_{\mathbf{n},k}\hat{E}^2_{\mathbf{n},k}-\frac{1}{2g^2}\sum_{\mathbf{n},ik}\left(\hat{U}^{\vphantom{\dagger}}_{\mathbf{n}\vphantom{+\hat{k}},i}\hat{U}^{\vphantom{\dagger}}_{\mathbf{n}+\hat{i},k}\hat{U}^\dagger_{\mathbf{n}+\hat{k},i}\hat{U}^\dagger_{\mathbf{n}\vphantom{+\hat{k}},k}+h.c.\right).
\end{equation}
By making $\epsilon^\prime=0$, the effective Hamiltonian simplifies significantly. Most of the correction terms disappear, and we are left with
\begin{equation}
\begin{aligned}
\hat{H}_{\mathrm{eff}}=&\left[\mu\left(1-\frac{\mu\epsilon^2}{\lambda^3}\right)+\frac{\mu\epsilon^2}{\lambda^2}\frac{1}{N_{0,l}(N_{0,l}+2)}\left(-2+\frac{9\mu}{\lambda}\right)+\frac{\epsilon^2}{\lambda}\frac{4}{N_{0,l}(N_{0,l}+2)}\left(1+\frac{18}{4}\frac{\epsilon^2}{\lambda^2}\right)\right] \sum_{\mathbf{n}} \hat{E}^2_{\mathbf{n},k}\\
&-\frac{5}{2}\frac{\epsilon^4}{\lambda^3}\sum_{\mathbf{n},ik}\left(\hat{\mathcal{U}}^{\vphantom{\dagger}}_{\mathbf{n}\vphantom{+\hat{k}},i}\hat{\mathcal{U}}^{\vphantom{\dagger}}_{\mathbf{n}+\hat{i},k}\hat{\mathcal{U}}^\dagger_{\mathbf{n}+\hat{k},i}\hat{\mathcal{U}}^\dagger_{\mathbf{n}\vphantom{+\hat{k}},k}+h.c.\right)+\frac{\epsilon^2\mu^2}{\lambda^3}\frac{6}{N_{0,l}(N_{0,l}+2)}\sum_{\mathbf{n}} \hat{E}_{\mathbf{n},k}\\
&+\frac{\epsilon^2\mu^2}{\lambda^3}\frac{8}{N_{0,l}(N_{0,l}+2)}\sum_{\mathbf{n}} \hat{E}^3_{\mathbf{n},k}+\frac{\epsilon^4}{\lambda^3}\frac{4}{N_{0,l}(N_{0,l}+2)}\left(\frac{\mu^2}{\epsilon^2}+\frac{2}{N_{0,l}(N_{0,l}+2)}\right)\sum_{\mathbf{n}} \hat{E}^4_{\mathbf{n},k}\\
&-\frac{\epsilon^4}{\lambda^3}\frac{2}{3}\frac{1}{\left(N_{0,l}(N_{0,l}+2)\right)^2}\sum_{\mathrm{``n.n."}}\left[\hat{E}_{\mathbf{n},k}\hat{E}_{\mathbf{n}^\prime,k^\prime}-11\hat{E}^2_{\mathbf{n},k}\hat{E}^2_{\mathbf{n}^\prime,k^\prime}\right]+\mathcal{O}\left(\frac{\epsilon^6}{\lambda^5}\right).
\end{aligned}
\end{equation}
This Hamiltonian, although simplified, still contains some non-desired correction terms. We study now how we can control them experimentally.

\subsubsection{Large Number of Atoms}

First, consider the limit where a very large number of atoms is placed on each link of the lattice, $N_{0,l}\rightarrow\infty$. Every correction term, except for a fourth order renormalization of the electric field, disappears in this limit. The non-unitary operators (\ref{eq:non_unitary}) converge to the correct unitary ones, and the effective Hamiltonian becomes the Kogut-Susskind Hamiltonian (\ref{eq:kogut_susskind_unitaries}),
\begin{equation}
\begin{aligned}
&H_{\mathrm{eff}}=\mu\left(1-\frac{\mu\epsilon^2}{\lambda^3}\right)\sum_{\mathbf{n}} \hat{E}^2_{\mathbf{n},k}-\frac{5}{2}\frac{\epsilon^4}{\lambda^3}\sum_{\mathbf{n},ik}\left(\hat{U}^{\vphantom{\dagger}}_{\mathbf{n}\vphantom{+\hat{k}},i}\hat{U}^{\vphantom{\dagger}}_{\mathbf{n}+\hat{i},k}\hat{U}^\dagger_{\mathbf{n}+\hat{k},i}\hat{U}^\dagger_{\mathbf{n}\vphantom{+\hat{k}},k}+h.c.\right),
\end{aligned}
\end{equation}
up to higher order corrections. These are, in the next orders, proportional to $\mu\epsilon^4/\lambda^4$, $\mu^3\epsilon^2/\lambda^4$ and $\epsilon^6/\lambda^5$. We will see that the last correction is the most relevant one. Note that, since no terms of order $\epsilon^{n}/\lambda^{n-1}$, for $n$ odd, appear in the effective expansion, it is enough to impose $\left(\epsilon/\lambda\right)^2\ll 1$ to make higher order correction negligible. This applies also to the rest of the situations we will consider. It is important to notice that, at the fourth order, the effective Hamiltonian in the large atom limit does not include any non-desired contribution, agreeing perfectly with the Kogut-Susskind Hamiltonian.

We write now the experimentally controllable parameters, $\epsilon$ and $\mu$, in terms of the coupling constant $g$ of the Kogut-Susskind Hamiltonian (\ref{eq:kogut_susskind_unitaries}). In order to do so, since we are only interested in the ratio between the electric and the magnetic part of the Hamiltonian, we multiply everything by a constant $\alpha$, whose effect is just to rescale the energy. This constant, and the coupling $g$, are defined as
\begin{equation}
\label{eq:g_limit}
\begin{aligned}
g&\equiv\left[\frac{2}{5}\frac{\mu\lambda^3}{\epsilon^4}\left(1-\frac{\mu\epsilon^2}{\lambda^3}\right)\right]^{1/4}, & \alpha&\equiv\left[\frac{1}{10}\frac{\lambda^3}{\mu\epsilon^4}\left(1-\frac{\mu\epsilon^2}{\lambda^3}\right)^{-1}\right]^{1/2}.
\end{aligned}
\end{equation}

The effective Hamiltonian can be expressed,then, as
\begin{equation}
\label{eq:KS_limit}
\begin{aligned}
\alpha H_{\mathrm{eff}}=&\frac{g^2}{2}\sum_{\mathbf{n}} \hat{E}^2_{\mathbf{n},k}-\frac{1}{2g^2}\sum_{\mathbf{n},ik}\left(\hat{U}^{\vphantom{\dagger}}_{\mathbf{n}\vphantom{+\hat{k}},i}\hat{U}^{\vphantom{\dagger}}_{\mathbf{n}+\hat{i},k}\hat{U}^\dagger_{\mathbf{n}+\hat{k},i}\hat{U}^\dagger_{\mathbf{n}\vphantom{+\hat{k}},k}+h.c.\right)\\
&+\mathcal{O}\left(\frac{1}{g^2}\frac{\mu}{\lambda}\left(\frac{\mu}{\epsilon}\right)^2\right)+\mathcal{O}\left(\frac{1}{g^2}\frac{\mu}{\lambda}\right)+\mathcal{O}\left(\frac{1}{g^2}\left(\frac{\epsilon}{\lambda}\right)^2\right),
\end{aligned}
\end{equation}
where the most important higher order corrections are written explicitly, allowing us to discuss their importance in the different regimes of the theory.

By using the Feshbach resonance technique to fine-tune the ratios $\frac{\epsilon}{\lambda}$ and $\frac{\mu}{\epsilon}$, one can explore the phase diagram of the Kogut-Susskind Hamiltonian. Consider, in particular, two extreme relevant limits.

\begin{itemize}
\item Strong coupling limit ($g\gg 1$)
\end{itemize}

To achieve this, we require
\begin{equation}
\left(\frac{\lambda}{\epsilon}\right)^3 \gg \frac{\epsilon}{\mu}\,.
\end{equation}
In this limit, only the electric part of the Hamiltonian is relevant. All the corrections are negligible as well.

\begin{itemize}
\item Weak coupling limit ($g\ll 1$)
\end{itemize}

In this case, we need
\begin{equation}
\label{eq:weak_limit}
\frac{\epsilon}{\mu} \gg \left(\frac{\lambda}{\epsilon}\right)^3.
\end{equation}

Here, the electric part vanishes and the plaquette interaction becomes relevant. The first two corrections in (\ref{eq:KS_limit}) are small compared to the latter, since $\epsilon/\mu$ is very large. Therefore, the biggest correction in this limit is of order $\left(\epsilon/\lambda\right)^2$ compared to the plaquette interaction.

From this, we conclude that both $\epsilon/\lambda$ and $\mu/\lambda$ must be small in order to make the higher order corrections in (\ref{eq:KS_limit}) negligible. However, we still have freedom to change the ratio $\epsilon/\mu$ to achieve different values of $g$.

\subsubsection{Finite Number of atoms}

Consider now a more realistic case, where the number of atoms on each link is finite, and non-desired contributions appear in the effective Hamiltonian. At the same time, a different approach to generate the electric field operators is introduced. We set $\mu=0$, and use the renormalization terms that were obtained in the effective expansion (at the second order) as the electric field part of the Hamiltonian. Proceeding this way, many correction terms disappear, and we are left with the following effective Hamiltonian,
\begin{equation}
\begin{aligned}
&H_{\mathrm{eff}}=\frac{\epsilon^2}{\lambda}\frac{4}{N_{0,l}(N_{0,l}+2)}\left(1+\frac{18}{4}\frac{\epsilon^2}{\lambda^2}\right) \sum_{\mathbf{n}} \hat{E}^2_{\mathbf{n},k}-\frac{5}{2}\frac{\epsilon^4}{\lambda^3}\sum_{\mathbf{n},ik}\left(\hat{\mathcal{U}}^{\vphantom{\dagger}}_{\mathbf{n}\vphantom{+\hat{k}},i}\hat{\mathcal{U}}^{\vphantom{\dagger}}_{\mathbf{n}+\hat{i},k}\hat{\mathcal{U}}^\dagger_{\mathbf{n}+\hat{k},i}\hat{\mathcal{U}}^\dagger_{\mathbf{n}\vphantom{+\hat{k}},k}+h.c.\right)\\
&+\frac{\epsilon^4}{\lambda^3}\frac{8}{\left(N_{0,l}(N_{0,l}+2)\right)^2}\sum_{\mathbf{n}} \hat{E}^4_{\mathbf{n},k}-\frac{\epsilon^4}{\lambda^3}\frac{2}{3}\frac{1}{\left(N_{0,l}(N_{0,l}+2)\right)^2}\sum_{\mathrm{``n.n."}}\left[\hat{E}_{\mathbf{n},k}\hat{E}_{\mathbf{n}^\prime,k^\prime}-11\hat{E}^2_{\mathbf{n},k}\hat{E}^2_{\mathbf{n}^\prime,k^\prime}\right]+\mathcal{O}\left(\frac{\epsilon^6}{\lambda^5}\right).
\end{aligned}
\end{equation}
Defining, as before, the coupling constant in terms of the system parameters in a proper way,
\begin{equation}
\label{eq:g}
g\equiv\left[\frac{8}{5N_{0,l}(N_{0,l}+2)}\frac{\lambda^2}{\epsilon^2}\left(1+\frac{18}{4}\frac{\epsilon^2}{\lambda^2}\right)\right]^{1/4},
\end{equation}
and multiplying everything by a constant
\begin{equation}
\label{eq:alpha}
\alpha\equiv\left[\frac{N_{0,l}(N_{0,l}+2)}{40}\frac{\lambda^4}{\epsilon^6}\left(1+\frac{18}{4}\frac{\epsilon^2}{\lambda^2}\right)^{-1}\right]^{1/2},
\end{equation}
the effective Hamiltonian can be written as
\begin{equation}
\label{eq:pure_gauge_corrections}
\begin{aligned}
\alpha \hat{H}_{\mathrm{eff}}=&\frac{g^2}{2}\sum_{\mathbf{n}} \hat{E}^2_{\mathbf{n},k}-\frac{1}{2g^2}\sum_{\mathbf{n},ik}\left(\hat{\mathcal{U}}^{\vphantom{\dagger}}_{\mathbf{n}\vphantom{+\hat{k}},i}\hat{\mathcal{U}}^{\vphantom{\dagger}}_{\mathbf{n}+\hat{i},k}\hat{\mathcal{U}}^\dagger_{\mathbf{n}+\hat{k},i}\hat{\mathcal{U}}^\dagger_{\mathbf{n}\vphantom{+\hat{k}},k}+h.c.\right)+\frac{8}{5g^2}\frac{1}{\left(N_{0,l}(N_{0,l}+2)\right)^2}\sum_{\mathbf{n}}\hat{E}^4_{\mathbf{n},k}\\
&-\frac{2}{15g^2}\frac{1}{\left(N_{0,l}(N_{0,l}+2)\right)^2}\sum_{\mathrm{``n.n."}}\left[\hat{E}_{\mathbf{n},k}\hat{E}_{\mathbf{n}^\prime,k^\prime}-11\hat{E}^2_{\mathbf{n},k}\hat{E}^2_{\mathbf{n}^\prime,k^\prime}\right]+\mathcal{O}\left(\frac{1}{g^2}\frac{\epsilon^2}{\lambda^2}\right).
\end{aligned}
\end{equation}
Since $N_{0,l}$ is a fixed quantity, once the experiment starts, the coupling $g$ is controlled with the ratio $\left(\epsilon/\lambda\right)^2$. Consider, again, the weak and strong coupling limits.

\begin{itemize}
\item Strong coupling limit ($g\gg 1$)
\end{itemize}

To access this regime, the experimental parameters must be chosen such that
\begin{equation}
\left(\frac{\lambda}{\epsilon}\right)^2 \gg N_{0,l}^2\,.
\end{equation}
Higher order corrections will be negligible provided that $\epsilon/\lambda\ll 1$. In this limit, the electric part of the Hamiltonian dominates. Therefore, even for a small number of atoms in each link, the atomic Hamiltonian accurately simulates the Kogut-Susskind one.

\begin{itemize}
\item Weak coupling limit ($g\ll 1$)
\end{itemize}

This regime is more problematic, and the accuracy of the simulation depends strongly on the number of atoms used in the experiment. To access it, we have to impose
\begin{equation}
\label{eq:weak_condition}
N_{0,l}^2 \gg \left(\frac{\lambda}{\epsilon}\right)^2 \gg 1\,,
\end{equation}
where the second condition is required to make higher order corrections negligible. In this regime, the number of atoms (and, correspondingly, the eigenvalues of the electric field operator) are not good quantum numbers, since the corresponding operators undergo large fluctuations. But as long as condition (\ref{eq:weak_condition}) is satisfied, the non-unitary operators are close enough to $\hat{U}_{\mathbf{n},k}=e^{-i\hat{\theta}_{\mathbf{n},k}}$, with $\hat{\theta}_{\mathbf{n},k}$ a well-defined hermitian phase operator. The latter provides good quantum numbers when large fluctuations are present.

In this regime, the non-desire fourth-order terms become important. All of them are of the order $\mathcal{O}\left(N^{-4}_{0,l}\right)$, but $E_{n,k}$ can take values $m$, depending on the state, from $-N_{0,l}/2$ to $N_{0,l}/2$. Higher values of $|m|$ are relevant when the plaquette interaction dominates over the electric part, however, their influence rapidly decreases with $N_{0,l}$. In particular, for a large number of atoms, the non-desired corrections vanish ``faster" than $\hat{\mathcal{U}}_{\mathbf{n},k}$ converges to $\hat{U}_{\mathbf{n},k}$, being the latter problem again the most important one.

A final comment about the energy scale of the simulated Hamiltonian is in order. To be able to obtain a plaquette interaction term that dominates over the electric field one, the latter should be made smaller. Therefore, the energy scale of the simulated Hamiltonian in the weak limit is very small. As a consequence, one would have to wait longer times to be able to observe relevant physical phenomena. In this situation, the coherence time of the experimental system plays an important role.

Apart from the two extreme limits, it is specially interesting to consider the intermediate case, with $g\sim 1$. This corresponds to the condition
\begin{equation}
N_{0,l}^2 \sim \left(\frac{\lambda}{\epsilon}\right)^2 \gg 1\,.
\end{equation}
This situation is not as hard to achieve in an experiment as the weak limit, but it is very relevant since most of the usual analytical approaches to study lattice gauge theories fail in this intermediate region \cite{Kogut-Susskind}.

Setting $\mu=0$ has several advantage with respect to the more conventional approach \cite{Erez_5}. As we have seen, it eliminates some of the non-desired correction terms. Also, the condition to enter the weak regime is easier to realize here, compared to (\ref{eq:weak_limit}). In this case, we can access the regime by making $\lambda/\epsilon$ smaller than $N_{0,l}$. The ratio should still be larger than $1$, but just enough to make higher order corrections small.

\subsection{Complete Abelian-Higgs Theory}

Let us consider now the complete Abelian-Higgs theory. Following the discussion of the pure-gauge case, we choose, again, $\mu=0$. This automatically removes many of the correction terms obtained in the last section.

Again, we multiply the effective Hamiltonian by a properly chosen constant $\alpha$, and define $g$ as before (\ref{eq:g}). In order to get the correct coupling constanst in front of the charge term, and the interaction term between the dynamical charges and the electric field, we define $R$ as
\begin{equation}
\label{eq:R_g}
R\equiv\frac{1}{g}\sqrt{\frac{2}{5}\left(\frac{\lambda}{\epsilon}\right)^3}\sqrt{\frac{|\epsilon^\prime|}{\epsilon}}\,.
\end{equation}
Once $g$ is fixed through $\lambda/\epsilon$, $R$ can be fine-tuned by changing the ratio $|\epsilon^\prime|/\epsilon$. Note that $\epsilon^\prime$ must be negative, since the ``charge" term and the gauge-matter interaction have opposite signs in the simulated Hamiltonian (\ref{eq:my_Hamiltonian}).

The total effective Hamiltonian reduces, then, to
\begin{equation}
\label{eq:simulated_Hamiltonian}
\begin{aligned}
\alpha \hat{H}_{\mathrm{eff}}&=\frac{1}{2R^2}\sum_\mathbf{n}\hat{Q}^2_\mathbf{n}-\frac{R^2}{2}\sum_{\mathbf{n},k}\left[\hat{\Phi}^\dagger_{\mathbf{n}\vphantom{\hat{k}}}\hat{\mathcal{U}}^{\dagger}_{\mathbf{n}\vphantom{\hat{k}},k}\hat{\Phi}^{\vphantom{\dagger}}_{\mathbf{n}+\hat{k}}+h.c.\right]\\
&+\frac{g^2}{2}\sum_{\mathbf{n}} \hat{E}^2_{\mathbf{n},k}-\frac{1}{2g^2}\sum_{\mathbf{n},ik}\left(\hat{\mathcal{U}}^{\vphantom{\dagger}}_{\mathbf{n}\vphantom{+\hat{k}},i}\hat{\mathcal{U}}^{\vphantom{\dagger}}_{\mathbf{n}+\hat{i},k}\hat{\mathcal{U}}^\dagger_{\mathbf{n}+\hat{k},i}\hat{\mathcal{U}}^\dagger_{\mathbf{n}\vphantom{+\hat{k}},k}+h.c.\right)\\
&-\frac{2}{15g^2}\frac{1}{\left(N_{0,l}(N_{0,l}+2)\right)^2}\sum_{\mathrm{``n.n."}}\left[\hat{E}_{\mathbf{n},k}\hat{E}_{\mathbf{n}^\prime,k^\prime}-11\hat{E}^2_{\mathbf{n},k}\hat{E}^2_{\mathbf{n}^\prime,k^\prime}\right]\\
&+\frac{8}{5g^2}\frac{1}{\left(N_{0,l}(N_{0,l}+2)\right)^2}\sum_{\mathbf{n}}\hat{E}^4_{\mathbf{n},k}+\alpha\hat{H}^\prime_\mathrm{eff}+\mathcal{O}\left(\frac{\epsilon^2}{\lambda^2}\frac{1}{g^2}\right)
\end{aligned}
\end{equation}
up to higher order effective contributions, which are negligible for all values of $g$ provided that the conditions introduced in the pure-gauge case are fulfilled. $\hat{H}^\prime_\mathrm{eff}$ comprises the non-desired effective corrections (up to fourth order) that are small enough to neglect (\ref{app:negligible}).

The phase diagram of the theory can be explored by changing the ratios $\frac{\epsilon}{\lambda}$ and  $\frac{|\epsilon^\prime|}{\epsilon}$, at the same time as we maintain most of non-desired terms small enough. For small values of $R$ and large values of $g$, the atomic system maps exactly to the Abelian-Higgs theory. When $R$ grows and $g$ decreases, the effect of the non-unitary character of the operators $\hat{\mathcal{U}}_{\mathbf{n},k}$ and $\hat{\Phi}_{\mathbf{n}}$ becomes important, as well as the relevance of the non-desired electric field terms, as it was discussed for the pure-gauge case. Making $N_{0,l}$ and $N_{0,v}$ larger will allow us to perform better simulations, going deeper into the large $R$ and small $g$ regime. The discussion for the strong regime in $R$ is analogous to the weak regime in $g$, and vice versa.

\section{Possible Experiments}
\label{sec:possible_experiments}

In the last section, we derived the experimental conditions under which the Hamiltonian describing a system of ultracold bosons is well approximated by the Hamiltonian associated to the Abelian-Higgs lattice gauge theory (\ref{eq:my_Hamiltonian}). Once this is achieved, several experiments can be perform on such a system, providing useful information about different high energy physics phenomena.

\subsection{Initialization: Static Charge Sectors}

\begin{figure}[t]
\centering
\includegraphics[width=0.8\textwidth]{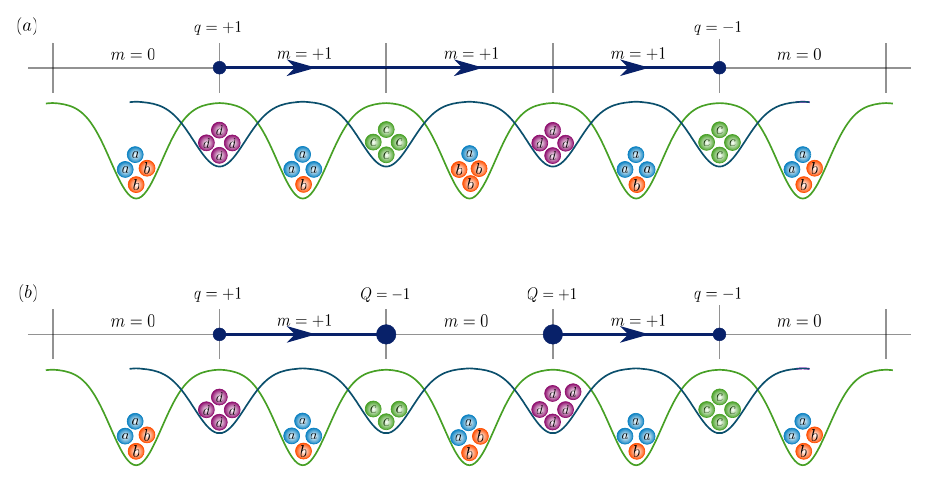}
\caption{In the figure, the state of both the simulating and the simulated systems is represented in two different situations. Figure (a) corresponds to the state of the system at the beginning of the simulation, with $R=0$ and $g\gg 1$. In this situation, the ground state of the simulated system, in a sector characterized by two static charges of opposite sign ($q=+1$ and $q=-1$), corresponds to an electric flux line ($m=+1$) between the static charges, and zero dynamical charges. The latter is achieved by placing $N_{0,v}=4$ bosons on each vertex. The electric field on the links, on the other hand, depends on the difference between the number of atoms of species $a$ and $b$, which interchange their role in alternating links. When $R$ is increased, the state represented in (a) is no longer an eigenstate of the Hamiltonian. The electric flux line can break (b), along with the appearance of two dynamical charges ($Q=+1$ and $Q=-1$), leaving the system in a superposition of the states (a) and (b)\textemdash notice that both of them fulfill the corresponding Gauss law at each vertex (\ref{eq:gauss_law}). The dynamical breaking of electric flux lines can be observed, in real time, by measuring the state of the ultracold atomic system.}
\label{fig:measurements}
\end{figure}

As a first step, the atomic system is initialized in the ground state of the non-interacting part of the Hamiltonian (\ref{eq:my_Hamiltonian}). Therefore, we choose the experimental parameters such that $g\gg 1$ (\ref{eq:g}) and $R=0$ (\ref{eq:R_g}). Consider the vertices of the lattice. Any dynamical charge configuration maps directly to the number of dynamical bosons of species $c$ and $d$ (\ref{eq:charge_operators}). The ground state corresponds, in particular, to an initial atomic system with $N_{0,v}$ dynamical bosons on each vertex (of the corresponding species, with the same number for each case), such that there are no dynamical charges present at the beginning of the simulation (Fig. \ref{fig:measurements}a). In addition, the vertices of the lattice are also filled with one auxiliary boson per site, alternating the two species ($e$ and $f$) on odd and even vertices, as we described in Section \ref{sec:effective}.

For the links, there is also a mapping between the eigenstates of the electric part of the Hamiltonian and the difference in the number of gauge bosons of species $a$ and $b$ (\ref{eq:electric_field_difference}), noting that their roles interchange (\ref{eq:canonical_transformation}). The difference here is that the ground state of the Hamiltonian does not correspond, in general, to the product of the ground states of each link (zero electric field configuration). As a consequence of gauge invariance, the Hilbert space of the system is divided into different sectors (\ref{eq:sectors})\textemdash characterized by different static charge configurations\textemdash and the states that belong to them are constrained by the Gauss law (\ref{eq:gauss_law}).

Consider, for example, the sector characterized by two static charges of opposite sign (Fig. \ref{fig:measurements}a). The states that belong to such a sector present an electric flux line\textemdash or a superposition of them\textemdash that joins the two vertices where the static charges are placed. Consequently, some of the links will be in excited states of the electric part of the Hamiltonian, whereas others remain in the ground state\textemdash since, altogether, everything must satisfy all the local Gauss laws (\ref{eq:gauss_law}). The ground state corresponds to the shortest path of electric field excitations between the charges, in this case, a straight line. This is the situation, in general, for any gauge-invariant sector. Therefore, any configuration of static charges can be achieved by initializing the atomic system with the proper number of atoms of type $a$ and $b$. Since this number will not be the same on every link for a non-zero static charge configuration (as opposed to the vertices), single-atom addressing techniques will be required \cite{single_atom_adressing_2}.

\subsection{Turning on the Interactions}

Once the system is initialized in the ground state of the non-interacting part of the Hamiltonian, the interactions can be introduced by modifying the value of the coupling constants $R$ and $g$, tuning the corresponding experimental parameters (\ref{eq:R_g}). This can be done either adiabatically or suddenly (quench). If the change is adiabatic, the system will adapt to a new state that corresponds to the ground state of the new Hamiltonian. Measuring the new state allows us to explore the phase diagram of the system. In the case of a quench, the system will not have enough time to adapt to the change in the Hamiltonian. Instead, it will remain in the initial state, which will correspond now to an excited state of the new Hamiltonian. Here, measuring the state of the system will provide us with information about the non-equilibrium and thermalization properties of the theory.

\subsection{Real-Time Dynamics}

We propose two different dynamical effects that can be measured in this setup.

\subsubsection{Fluctuation of Electric Flux Lines}

A initial state with well-defined electric flux lines is an eigenstate of the non-interacting part of the Hamiltonian. Reducing the value of $g$ increases the probability of a plaquette-type electric flux loop to appear (Fig. \ref{fig:static_charges}). If we change this parameter adiabatically, more of these flux loops will appear, being the state of the system a superposition of many different electric field configurations. The initial straight flux line can be deformed if one plaquette excitation coincides with one of its links (Fig. \ref{fig:static_charges}). This effect can be detected dynamically by measuring the state of the atoms on the links of the lattice.

\subsubsection{Breaking of Electric Flux Lines}

Another effect that could be measured dynamically is the breaking of electric flux lines (Fig. \ref{fig:measurements}). By increasing the value of $R$, the probability of new pairs of dynamical charges of opposite sign to appear grows. If such excitations coincide with an existing electric field line between other pair of charges, it will cause the breaking of the latter, resulting in two new pairs joined by shorter lines (Fig. \ref{fig:measurements}).

This phenomenon, relevant in the understanding of confinement of matter in gauge theories, is hard to study using analytical or numerical techniques, since real-time dynamics are difficult to simulate using standard Monte Carlo methods. Using this quantum simulation scheme, the dynamical breaking of electric fluxes could be accurately measured.

\subsection{Measuring the State of the System}

We showed that there is a mapping between the state of the simulated system in the basis of the eigenstate of the electric field and charge operator, and the number of atoms of each species on the vertices and links of the lattice. Therefore, in order to know the state of the system at any point of the simulation, it is enough to measure the number of atoms locally, using single-atom detection techniques \cite{single_atom_adressing_1}. By doing that we could obtain a detail picture of the flux lines and dynamical charge configuration that characterizes the state of the system.

\section{Summary}
\label{sec:conclusions}

In this paper, we proposed a quantum simulation scheme for the lattice version of the Abelian-Higgs theory, using ultracold bosonic atoms trapped in an optical lattice. The interest in performing such simulation is twofold. On the one hand, and despite its simplicity, this model shows very interesting high energy physics phenomena. It allows the possibility to study both the Brout-Englert-Higgs mechanism and the confinement of dynamical Higgs matter. On the other hand, it can serve as a benchmark to study the validity of several quantum simulation techniques, since the phase diagram of this theory is well known from conventional theoretical and numerical calculations, and can be compared to the one obtained in a quantum simulation.

Starting from the lattice action of the Abelian-Higgs theory \cite{Fradkin_Shenker}, the corresponding quantum Hamiltonian was obtained using the transfer matrix method (\ref{eq:my_Hamiltonian}). Such a Hamiltonian is a generalization of the pure-gauge Hamiltonian obtained by Kogut and Susskind \cite{Kogut-Susskind}, including, as well, the degrees of freedom corresponding to the scalar matter fields.

The quantum simulation of this model using ultracold atoms is achieved by utilizing two superimposed optical lattices, filled with six different bosonic species. These correspond to three different types of atoms, each one possessing two accessible hyperfine angular momentum levels. One type of atoms simulates the gauge degrees of freedom, and another one the dynamical matter. The last type of atoms serves as auxiliary particles, required for obtaining the plaquette interactions. An intermediate ``primitive Hamiltonian" is first obtained by selecting the correct hyperfine angular momentum levels for the six species, and tuning some atomic scattering lengths using Feshbach resonances. After transforming the atomic degrees of freedom and expressing them in terms of the lattice Abelian-Higgs operators, the atomic Hamiltonian results in a gauge-invariant Hamiltonian. The corresponding operators are approximated with the bosonic ones, improving the approximation when the number of bosons increases. The primitive Hamiltonian contains all the terms of the Abelian-Higgs Hamiltonian except for the plaquette interactions. From the former, and with the help of the ``hard-core" auxiliary bosons, the Abelian-Higgs Hamiltonian can be approximated effectively, provided that the energy scales of the system are tuned correctly. The degree of approximation can be controlled by increasing the number of atoms used in the simulation, making the non-desired correction terms in the effective expansion negligible.

Finally, we described an experimental scheme for measuring, in real time, interesting phenomena such as the fluctuation and breaking of electric flux lines, and for adiabatically preparing the system in the ground state, which could help to explore the phase diagram of this high energy physics theory.

Next steps toward a complete understanding of the Abelian-Higgs theory might include higher dimensional cases, as well as non-fundamental representations of the gauge group ($q>1$), allowing a richer phase diagram to be accessed in a quantum simulation experiment. In a more general context, scalar Higgs matter could be also studied for non-abelian theories.

\section{Acknowledgements}

This project has received funding from the European Union's Horizon 2020 research and innovation programme under the Marie Sk\l{}odowska-Curie grant agreement No 665884, the Spanish Ministry of Economy and Competitiveness (SEVERO OCHOA Grants No. SEV-2015-0522 and No. FISICATEAMO FIS2016-79508-P), Agencia de Gesti\`{o} d'Ajuts Universitaris i de Recerca (2014SGR874 and CERCA Program) and Fundaci\'{o} Cellex.

\appendix

\section{The Abelian Higgs Model}
\label{app:AbHiggs}

\subsection{Review of the Brout-Englert-Higgs Mechanism}
\label{app:higgs}
Consider the following Lagrangian density,
\begin{equation}
\label{eq:abelian_higgs_lagrangian}
\begin{aligned}
\mathcal{L}=(\partial_\mu\phi^*+igA_\mu\phi^*)(\partial_\mu\phi-igA_\mu\phi)+m^2\phi^*\phi-\frac{\lambda}{2}(\phi^*\phi)^2-\frac{1}{4}F_{\mu\nu}F^{\mu\nu},
\end{aligned}
\end{equation}
where $\phi(x)$ is a complex scalar field, $A_\mu(x)$ is a massless vector (gauge) field\textemdash which we will call the photonic field\textemdash and $g$ is the interaction strength (charge). Finally, $F_{\mu\nu}$ is the usual electromagnetic field tensor,
\begin{equation}
\label{eq:em_tensor}
F_{\mu\nu}(x)=\partial_\mu A_\nu(x)-\partial_\nu A_\mu(x)\,.
\end{equation}

This theory is invariant under the group of local $U(1)$ transformations,
\begin{equation}
\begin{aligned}
\phi(x)&\rightarrow e^{i\alpha(x)} \phi(x), &
A_\mu(x)&\rightarrow A_\mu(x)-\frac{1}{g}\partial_\mu \alpha(x)\,,
\end{aligned}
\end{equation}
parametrized by $\alpha(x)$, which can be different at each space-time point, locally.

The Lagrangian also includes a quartic potential for the scalar field, the \textit{Higgs potential},
\begin{equation}
V(\phi)=-m^2|\phi|^2+\frac{\lambda}{2}|\phi|^4\,.
\end{equation}
A semiclassical approximation for the expectation value of the field $\phi$ in the ground state is obtained by minimizing the above potential, obtaining a non-zero value for $m^2>0$,
\begin{equation}
\ex{\phi}_0=\sqrt{\frac{2m^2}{\lambda}}\equiv\frac{R}{\sqrt{2}}\,.
\end{equation}
For $m^2<0$, however, the minimum is found for $\ex{\phi}_0=0$.

To take into account the effect of quantum fluctuations, one can expand the complex field $\phi(x)$ around the semiclassical result, using two real scalar fields, $\sigma(x)$ and $\pi(x)$, such that
\begin{equation}
\phi(x)=\left(\frac{R+\sigma(x)}{\sqrt{2}}\right)e^{i\pi(x)/R}.
\end{equation}
Plugging this expression into (\ref{eq:abelian_higgs_lagrangian}), the Lagrangian can be expressed in terms of the new fields. In particular, the terms containing the gauge field $A_\mu$ alone are now the following,
\begin{equation}
\mathcal{L}=-\frac{1}{4}F_{\mu\nu}F^{\mu\nu}+\frac{1}{2}m^2_A A^\mu A_\mu+...\,,
\end{equation}
with $m_A\equiv gR$. Thus, the non-zero expectation value of the scalar field $\phi$ generates a mass term for the gauge field $A_\mu$ or, in other words, the photon becomes massive, which shows how the Higgs mechanism takes place in this simple model.

In the limit $m,\lambda\rightarrow\infty$, with $m^2/\lambda$ fixed, the radial field $\sigma$ decouples, and the effective Lagrangian for the remaining fields is just
\begin{equation}
\label{eq:effective_lagrangian}
\mathcal{L}_\mathrm{eff}=\frac{1}{2}\partial^\mu\pi\partial_\mu\pi+\frac{1}{2}m_A^2A^\mu A_\mu-m_AA^\mu\partial_\mu\pi-\frac{1}{4}F_{\mu\nu}F^{\mu\nu}.
\end{equation}

\subsection{Lattice Gauge Action}
\label{app:action}

The Abelian-Higgs model can be formulated as a lattice gauge theory \cite{Fradkin_Shenker,Einhorn_Savit_1,Einhorn_Savit_2,Jones_Kogut_Sinclair,Ostewalder_Seiler,Abelian_Higgs_Montecarlo,abelian_higgs_4}, this is, as a field theory on a discretized space-time. The Euclidean lattice action corresponding to the effective theory (\ref{eq:effective_lagrangian}) has the following form, after a Wick rotation is applied \cite{Fradkin_Shenker},
\begin{equation}
\label{eq:lattice_gauge_action}
\begin{aligned}
S_\mathrm{E}=-\frac{R^2a^{d-1}}{2}\sum_{n,\mu}\left[e^{i\varphi_n}e^{iq\theta_{n,\mu}}e^{-i\varphi_{n+\hat{\mu}}}+h.c.\right]-\frac{a^{d-3}}{2g^2}\sum_{n,\mu,\nu}\left(e^{i\theta_{n,\mu}}e^{i\theta_{n+\hat{\mu},\nu}}e^{-i\theta_{n+\hat{\nu},\mu}}e^{-i\theta_{n,\nu}}+h.c\right),
\end{aligned}
\end{equation}
 where the fields $\phi_n\equiv e^{-i\varphi_n}$ reside on the vertices of a $d+1$ dimensional square lattice with lattice spacing $a$, whereas the fields $U^{(q)}_{n,\mu}\equiv e^{-iq\theta_{n,\mu}}$ reside on the links that joins the vertices $n$ and $n+a\,\hat{\mu}$, with $\hat{\mu}$ a unit vector in one of the $d+1$ orthogonal directions. The coupling constants $R$ and $g$ are introduced to obtain the correct expression in the continuum limit. Finally, $q$ is an integer number denoting the representation of the $U(1)$ group under which the Higgs field is transforming, where the fundamental one corresponds to $q=1$.

The first term in (\ref{eq:lattice_gauge_action}) is an interaction between nearest-neighbor vertices, mediated by the link that connects them. The second one is a \textit{plaquette interaction}, which involves the fields on the four links of a plaquette.

The phase diagram of this lattice gauge theory depends on the specific group representation under which the Higgs field transforms. In particular, one finds a phase diagram with two phases if the representation is the fundamental one ($q=1$), and another one with three phases if any other representation is used \cite{Fradkin_Shenker,Abelian_Higgs_Montecarlo}. We shall describe now the properties of the possible phases one can find in the theory for $d+1\geq 4$, based on the results from \cite{Fradkin_Shenker}. If the Higgs field transforms according to a representation of $U(1)$ with $q>1$, the theory presents three different phases,

\begin{itemize}
\item Higgs phase (large $R$, small $g$). In this phase the photon is massive (Higgs mechanism), the force between different particles is short-ranged, and there is no confinement of charges.

\item Coulomb phase (small $R$, small $g$). Here, the photon is massless, and the charges are free, as well.

\item Confinement phase (large $g$). In this regime there are no free charges (confinement) and the photon has a finite mass.

\end{itemize}

In the fundamental representation ($q=1$), it can be shown that there is no phase boundary between the Higgs and confining regimes \cite{Fradkin_Shenker}. The theory presents only two different phases, a Coulomb phase as in the previous case, and a Higgs-confinement phase, characterized by massive photons, short-ranged forces, and the absence of free charges.

In the $2+1$ dimensional case there is no Coulomb phase for finite coupling constants. In the fundamental representation, there is only one phase, the Higgs-confinement one, whereas for $q>1$, there is a phase boundary separating the Higgs and confinement regimes \cite{Fradkin_Shenker}.

\section{Hamiltonian Formulation}
\label{sec:hamiltonian_formulation}

We compute the quantum Hamiltonian corresponding to the Abelian-Higgs action (\ref{eq:lattice_gauge_action}) by applying the transfer matrix method. The application of the method to the pure-gauge theory leads to the standard Kogut-Susskind Hamiltonian \cite{Kogut}. Here, we extend it to the matter part of the action for the fundamental representation of $U(1)$ ($q=1$),
\begin{equation}
S=-\frac{R^2a^{d-1}}{2}\sum_{n,\mu}\left[e^{i\varphi_n}e^{i\theta_{n,\mu}}e^{-i\varphi_{n+\hat{\mu}}}+h.c.\right].
\end{equation}

First, we separate the temporal and the spatial directions. We introduce the notation $\mu=(0,k)$, with $k=1,...,d$, and $n=(n_0,\mathbf{n})$, where the first component corresponds to the temporal direction and the second to the $d$ spatial ones. Both $n_0$ and $\mathbf{n}$ take integer values from $0$ to $N$. We also consider the \textit{temporal gauge},
\begin{equation}
\theta_{n,0}=0\,.
\end{equation}
The action is expressed, then, as
\begin{equation}
\begin{aligned}
&S=-\frac{R^2a^d}{a_\tau}\sum_{\mathbf{n},n_0} \mathrm{cos}(\varphi_{\mathbf{n},n_0+1}-\varphi_{\mathbf{n},n_0\vphantom{\hat{0}}})-R^2a^{d-2}a_\tau\sum_{\mathbf{n},k,n_0}\mathrm{cos}(\varphi_{\mathbf{n}+\hat{k},n_0}-\varphi_{\mathbf{n}\vphantom{\hat{k}},n_0}-\theta_{\mathbf{n}\vphantom{\hat{k}},k,n_0})\,,
\end{aligned}
\end{equation}
where $a_\tau$ and $a$ are the lattice spacings in the temporal and spatial directions, respectively. To obtain the correct continuum limit, the fields $\varphi_n$ should depend on $a_\tau$ as $\varphi_{\mathbf{n},n_0+1}-\varphi_{\mathbf{n},n_0\vphantom{\hat{0}}}\propto a_\tau$. Therefore, we can expand the cosine in terms of this parameter. We get, disregarding irrelevant terms (in the limit $a_\tau\rightarrow 0$),
\begin{equation}
\begin{aligned}
&S=\frac{R^2a^d}{2a_\tau}\sum_{\mathbf{n},n_0}(\varphi_{\mathbf{n},n_0+1}-\varphi_{\mathbf{n},n_0\vphantom{\hat{0}}})^2-R^2a^{d-2}a_\tau\sum_{\mathbf{n},k,n_0}\mathrm{cos}(\varphi_{\mathbf{n}+\hat{k},n_0}-\varphi_{\mathbf{n}\vphantom{\hat{k}},n_0}-\theta_{\mathbf{n}\vphantom{\hat{k}},k,n_0})\,.
\end{aligned}
\end{equation}

In order to connect the path integral and the Hamiltonian formulations of quantum mechanics, it is useful to consider the partition function,
\begin{equation}
\label{eq:partition_function}
Z=\int\!D\tilde{\varphi}\, e^{-S}=\int\!D\tilde{\varphi}\, \prod_{n_0}e^{-S(\varphi_{n_0+1},\varphi_{n_0})},
\end{equation}
where $\varphi_{n_0}$ denotes the set of fields variables $\varphi_{\mathbf{n},n_0}$ for all $\mathbf{n}$, and $\tilde{\varphi}$ refers to the complete set of $\varphi_{n_0}$, for all $n_0$. In the previous expression, the action was split into a sum of different terms,
\begin{equation}
S=\sum_{n_0}S(\varphi_{n_0+1},\varphi_{n_0})\,,
\end{equation}
with
\begin{equation}
\begin{aligned}
S(\varphi_{n_0+1},\varphi_{n_0})&=\frac{R^2a^d}{2a_\tau}\sum_{\mathbf{n}}(\varphi_{\mathbf{n},n_0+1}-\varphi_{\mathbf{n},n_0\vphantom{\hat{0}}})^2-\frac{R^2a^{d-2}a_\tau}{2}\sum_{\mathbf{n},k}\mathrm{cos}(\varphi_{\mathbf{n}+\hat{k},n_0}-\varphi_{\mathbf{n}\vphantom{\hat{k}},n_0}-\theta_{\mathbf{n}\vphantom{\hat{k}},k,n_0})\\
&-\frac{R^2a^{d-2}a_\tau}{2}\sum_{\mathbf{n},k}\mathrm{cos}(\varphi_{\mathbf{n}+\hat{k},n_0+1}-\varphi_{\mathbf{n}\vphantom{\hat{k}},n_0+1}-\theta_{\mathbf{n}\vphantom{\hat{k}},k,n_0+1})\,.
\end{aligned}
\end{equation}

The \textit{transfer matrix elements} are defined as $T(\varphi^\prime,\varphi)=e^{-S(\varphi^\prime,\varphi)}$, and can be thought as the matrix elements of a \textit{transfer operator} $\hat{T}$ over a complete and orthonormal product basis $\{\ket{\ \varphi}=\prod_{\mathrm{n}}\ket{\varphi_{\mathrm{n}}}\}$,
\begin{equation}
\bra{\varphi^\prime}\hat{T}\ket{\varphi^{\vphantom{\prime}}}=T(\varphi^\prime,\varphi)\,.
\end{equation}

The partition function is expressed in terms of the transfer operator as
\begin{equation}
\begin{aligned}
Z&=\int\!D\tilde{\varphi}\,\prod_{n_0}\bra{\varphi_{n_0+1}}\hat{T}\ket{\varphi_{n_0}}=\int\!D\varphi_0 D\varphi_{N+1}\,\bra{\varphi_{N+1}}\hat{T}^N\ket{\varphi_0}\,,
\end{aligned}
\end{equation}
where, in the last equality, the completion relation for the basis,
\begin{equation}
\mathbb{I}=\int\!D\varphi\,\ket{\varphi}\bra{\varphi}\,,
\end{equation}
was used $N-1$ times. Using periodic boundary conditions, $\varphi_0=\varphi_{N}\equiv\varphi$, the partition function can be expressed as a trace,
\begin{equation}
Z=\int\!D\varphi\,\bra{\varphi}\hat{T}^N\ket{\varphi}=\tr{T^N}.
\end{equation}
Motivated by the latter, a quantum Hamiltonian is defined through the transfer operator,
\begin{equation}
\label{eq:T_H}
\hat{T}=\left(e^{-\tau \hat{H}}\right)^{1/N}\longrightarrow e^{-a_\tau\hat{H}},
\end{equation}
taking the continuum limit in the temporal direction, $N\rightarrow\infty$, $a_\tau\rightarrow 0$, with $\tau=N a_\tau$ fixed.

In order to get an expression for the transfer operator, we introduced the second-quantized operators $\hat{\varphi}_\mathrm{n}$ and $\hat{Q}_\mathrm{n}$, following the commutation relations
\begin{equation}
\label{eq:commutator_Q_Phi}
[\hat{\varphi}_\mathbf{n},\hat{Q}_{\mathbf{n}^\prime}]=i\delta_{\mathbf{n},\mathbf{n}^\prime}\,.
\end{equation}
$\hat{\varphi}_\mathbf{n}$ is defined such that
\begin{equation}
\hat{\varphi}_\mathbf{n}\ket{\varphi}=\varphi_\mathbf{n}\ket{\varphi}\,.
\end{equation}

It can be shown that the following property holds,
\begin{equation}
\bra{\varphi^\prime}\mathrm{exp}\left(-\frac{a^{-d}}{2R^2}a_\tau\hat{Q}^2_\mathbf{n}\right)\ket{\varphi}\propto \mathrm{exp} \left(-\frac{R^2a^d}{2}\frac{(\varphi^\prime_\mathbf{n}-\varphi^{\vphantom{\prime}}_\mathbf{n})^2}{a_\tau}\right),
\end{equation}
by using the commutation relations (\ref{eq:commutator_Q_Phi}). With this, it is easy to see that the transfer operator can be expressed as
\begin{equation}
\begin{aligned}
\hat{T}=&\mathrm{exp}\left[-a_\tau\frac{R^2a^{d-2}}{2}\sum_{\mathbf{n},k}\mathrm{cos}(\hat{\varphi}_{\mathbf{n}+\hat{k}}-\hat{\varphi}_{\mathbf{n}\vphantom{\hat{k}}}-\hat{\theta}_{\mathbf{n}\vphantom{\hat{k}},k})\right]\mathrm{exp}\left[-a_\tau\frac{a^{-d}}{2R^2}\sum_\mathbf{n}\hat{Q}^2_\mathbf{n}\right]\\
\times&\mathrm{exp}\left[-a_\tau\frac{R^2a^{d-2}}{2}\sum_{\mathbf{n},k}\mathrm{cos}(\hat{\varphi}_{\mathbf{n}+\hat{k}}-\hat{\varphi}_{\mathbf{n}\vphantom{\hat{k}}}-\hat{\theta}_{\mathbf{n}\vphantom{\hat{k}},k})\right],
\end{aligned}
\end{equation}
where $\hat{\theta}_{\mathbf{n}\vphantom{\hat{k}},k}$ is a quantum operator corresponding to the gauge field $\theta_{\mathbf{n}\vphantom{\hat{k}},k}$, obtained after applying the similar procedure to the gauge degrees of freedom on the links \cite{Kogut}.

Taking the continuum limit in the time direction, $a_\tau$, one can read the Hamiltonian (up to irrelevant constants) from the transfer operator, using equation (\ref{eq:T_H}),
\begin{equation}
\hat{H}=\frac{a^{-d}}{2R^2}\sum_\mathbf{n}\hat{Q}^2_\mathbf{n}-\frac{R^2a^{d-2}}{2}\sum_{\mathbf{n},k}\left[\phi^\dagger_{\mathbf{n}\vphantom{\hat{k}}}\hat{U}^\dagger_{\mathbf{n}\vphantom{\hat{k}},k}\hat{\phi}^{\vphantom{\dagger}}_{\mathbf{n}+\hat{k}}+h.c.\right],
\end{equation}
where $\hat{\phi}_\mathbf{n}\equiv e^{-i\hat{\varphi}_\mathbf{n}}$ and $\hat{U}_{\mathbf{n},k}\equiv e^{-i\hat{\theta}_{\mathbf{n},k}}$. In order to avoid the \textit{quantum phase operator} problem \cite{Phase_0,Phase_1,Phase_2,Phase_3,Phase_4}, the operator $\hat{Q}_\mathbf{n}$ must be unbounded from below. Only in that case $e^{i\hat{\varphi}_\mathbf{n}}$ is a well-defined unitary operator. The situation is equivalent for the conjugate momentum of $\hat{\theta}_{\mathbf{n},k}$, $\hat{E}_{\mathbf{n},k}$, defined in \cite{Kogut} to derive the pure-gauge part of the Hamiltonian via the commutation relations
\begin{equation}
[\hat{\theta}_{\mathbf{n},k},\hat{E}_{\mathbf{n}^\prime,k^\prime}]=i\delta_{k,k^\prime}\delta_{\mathbf{n},\mathbf{n}^\prime}\,.
\end{equation}

The complete Hamiltonian associated to the Abelian-Higgs theory, including the pure-gauge part \cite{Kogut}, is
\begin{equation}
\begin{aligned}
\hat{H}&=\frac{1}{2R^2}\sum_\mathbf{n}\hat{Q}^2_\mathbf{n}-\frac{R^2}{2}\sum_{\mathbf{n},k}\left[\phi^\dagger_{\mathbf{n}\vphantom{\hat{k}}}\hat{U}^\dagger_{\mathbf{n}\vphantom{\hat{k}},k}\hat{\phi}^{\vphantom{\dagger}}_{\mathbf{n}+\hat{k}}+h.c.\right]\\
&+\frac{g^2}{2}\sum_{\mathbf{n},k}\hat{E}^2_{\mathbf{n},k}-\frac{1}{2g^2}\sum_{\mathbf{n},ik}\left(\hat{U}^{\vphantom{\dagger}}_{\mathbf{n}\vphantom{+\hat{k}},i}\hat{U}^{\vphantom{\dagger}}_{\mathbf{n}+\hat{i},k}\hat{U}^\dagger_{\mathbf{n}+\hat{k},i}\hat{U}^\dagger_{\mathbf{n}\vphantom{+\hat{k}},k}+h.c.\right),
\end{aligned}
\end{equation}
where we set $a=1$.

\section{Non-Commuting Operators}
\label{app:non_commuting}

Let $\hat{A}$, $\hat{B}$ and $\hat{C}$ be operators, fulfilling the commutation relation
\begin{equation}
\label{eq:commuting_properties}
[\hat{A},\hat{B}]=\hat{C}\hat{B}\,,
\end{equation}
with $[\hat{A},\hat{C}]=0$ and $[\hat{B},\hat{C}]=0$. Then, for any analytical function $f$, acting on the operators through its Taylor series, the following property holds,
\begin{equation}
\label{eq:non_commuting_property}
f(\hat{A})\hat{B}=\hat{B}f(\hat{A}+\hat{C})\,.
\end{equation}
To see this, let us first prove that
\begin{equation}
\label{eq:lemma_n}
\hat{A}^n\hat{B}=B(\hat{A}+\hat{C})^n\,,
\end{equation}
where $n$ is any positive integer. We do this by induction. The case $n=1$ is (\ref{eq:commuting_properties}). Assume it is true for $n=k$. For $k+1$, we have
\begin{equation}
\hat{A}^{k+1}\hat{B}=\hat{A}^k\hat{B}(\hat{A}+\hat{C})=\hat{B}(\hat{A}+\hat{C})^{k+1}\,,
\end{equation}
where the property (\ref{eq:commuting_properties}) was used in the first equality, and the induction hypothesis in the second. Therefore, property (\ref{eq:lemma_n}) is true for $n>1$.

By Taylor expanding $f(\hat{A})$,
\begin{equation}
f(\hat{A})=\sum_n c_n \hat{A}^n\,,
\end{equation}
it is easy to obtain (\ref{eq:non_commuting_property}),
\begin{equation}
f(\hat{A})\hat{B}=\sum_n c_n \hat{A}^n\hat{B}=\hat{B}\sum_n c_n (\hat{A}+\hat{C})^n=\hat{B}f(\hat{A}+\hat{C})\,.
\end{equation}

Consider, as an example, the operators $\hat{E}$ and $\hat{U}^\dagger$, such that
\begin{equation}
[\hat{E},\hat{U}^\dagger]=\hat{U}^\dagger\,.
\end{equation}
In that case, $\hat{C}$ is the identity, and the property (\ref{eq:non_commuting_property}) reduces to
\begin{equation}
f(\hat{E})\hat{U}^\dagger=\hat{U}^\dagger f(\hat{E}+\mathbb{I})\,.
\end{equation}

\section{Effective Hamiltonian}
\label{app:calculations}

\subsection{Effective Expansion}
Here, details about the calculations that were carried out to obtain the effective Hamiltonian are provided. The property introduced in \ref{app:non_commuting} will be assumed henceforth. We follow the notation of \cite{eff_ham_1,eff_ham_2}. The starting point is the primitive Hamiltonian,
\begin{equation}
\label{eq:primitive_hamiltonian_app}
\begin{aligned}
&\hat{H}=\lambda \sum_{\mathbf{n}} \hat{N}^\chi_{\mathbf{n}}(\hat{N}^\chi_{\mathbf{n}}-1)+\epsilon \sum_{\mathbf{n},k} \left[\hat{\chi}^\dagger_{\mathbf{n}\vphantom{\hat{k}}}\hat{\mathcal{U}}^{\dagger}_{\mathbf{n}\vphantom{\hat{k}},k}\hat{\chi}^{\vphantom{\dagger}}_{\mathbf{n}+\hat{k}}+h.c.\right]\\
&+\mu \sum_{\mathbf{n},k} \hat{E}^2_{\mathbf{n},k}+\mu^\prime \sum_{\mathbf{n}} \hat{Q}^2_{\mathbf{n}}+\epsilon^\prime \sum_{\mathbf{n},k} \left[\hat{\Phi}^\dagger_{\mathbf{n}\vphantom{\hat{k}}}\hat{\mathcal{U}}^{\dagger}_{\mathbf{n}\vphantom{\hat{k}},k}\hat{\Phi}^{\vphantom{\dagger}}_{\mathbf{n}+\hat{k}}+h.c.\right].
\end{aligned}
\end{equation}

An effective Hamiltonian, acting on a subspace (sector) generated by an eigenvalue of the hard-core constraint
\begin{equation}
\hat{H}_\mathrm{HC}=\lambda \sum_{\mathbf{n}} \hat{N}^\chi_{\mathbf{n}}(\hat{N}^\chi_{\mathbf{n}}-1)
\end{equation}
is obtained perturbatively\textemdash provided that $\lambda$ is much larger than the rest of the energy scales.

The hard-core constraint is diagonal in the number basis of the auxiliary bosons, denoted by $\chi$. Its eigenvalues are largely degenerate, each of them associated with a set of distributions of auxiliary bosons over the vertices of the lattice. The ground state sector includes all the states characterized by either zero or one particle on each vertex. Our goal is to obtain an effective Hamiltonian that acts on the latter. The quantum simulator will be initialized, in particular, in a state with one auxiliary boson sitting at each vertex. In this situation, some of the fourth order corrections in the perturbative expansion of the effective Hamiltonian (Fig. \ref{fig:hopping_plaquette})  turn out to be the desired plaquette interactions of the Abelian-Higgs Hamiltonian (\ref{eq:my_Hamiltonian}). There are, of course, other contributions to the effective Hamiltonian. Here, we will analyze some of them. Instead of focusing on the detailed calculations, we will motivate the appearance of the correction terms by viewing them as virtual physical processes.

The primitive Hamiltonian (\ref{eq:primitive_hamiltonian_app}) contains two types of terms, apart from the hard-core constraint. First, we have those that do not commute with the latter. In particular, the hopping of auxiliary bosons to nearest-neighbor vertices, mediated by the gauge-field operators on the corresponding links, take the system out of the ground state sector. They form what we will call the interacting part of the Hamiltonian,
\begin{equation}
\hat{H}_\mathrm{int}=\epsilon \sum_{\mathbf{n},k}\left(\hat{\chi}^\dagger_{\mathbf{n}\vphantom{\hat{k}}}\hat{\mathcal{U}}^{\dagger}_{\mathbf{n}\vphantom{\hat{k}},k}\hat{\chi}^{\vphantom{\dagger}}_{\mathbf{n}+\hat{k}}+h.c.\right),
\end{equation}
where the notation introduced in (\ref{eq:non_unitary}) is assumed. The rest of the terms in (\ref{eq:primitive_hamiltonian_app}), on the other hand, keep the system inside the same sector.

The effective Hamiltonian will be composed of terms that commute with the hard-core constraint and, on top of them, higher order corrections made out of product of terms from $\hat{H}_\mathrm{int}$. The latter can be seen as non-trivial contributions resulting from virtual processes where the auxiliary bosons hop to nearest-neighbor vertices and then return to their original positions. Since the final state of the auxiliary bosons will not change, these degrees of freedom can be traced out of the effective Hamiltonian.

\subsubsection{First Order}

The first order contributions to the effective Hamiltonian are given by $P_0 H P_0$, where $P_0$ is the projection operator to the ground state sector $\mathcal{H}^0$,
\begin{equation}
\hat{P}_0=\sum_{\ket{\phi}\in\mathcal{H}^0}\ket{\phi}\bra{\phi}\,.
\end{equation}
In the case where the system is initialized in a state with one auxiliary boson per vertex, represented by $\ket{\phi_0}$, the projector reduces to $\ket{\phi_0}\bra{\phi_0}$, since the conservation of the total number of auxiliary particles prevents the system from being in other states from this sector\textemdash which have zero auxiliary bosons in some vertices, and one in the rest. At this order, only the terms that commute with the hard-core constraint contribute,
\begin{equation}
\begin{aligned}
&\hat{H}^{(1)}_{\mathrm{eff}}=\mu \sum_{\mathbf{n},k} \hat{E}^2_{\mathbf{n},k}+\mu^\prime \sum_{\mathbf{n}} \hat{Q}^2_{\mathbf{n}}+\epsilon^\prime \sum_{\mathbf{n},k} \left(\hat{\Phi}^\dagger_{\mathbf{n}\vphantom{\hat{k}}}\hat{\mathcal{U}}^{\dagger}_{\mathbf{n}\vphantom{\hat{k}},k}\hat{\Phi}^{\vphantom{\dagger}}_{\mathbf{n}+\hat{k}}+h.c.\right).
\end{aligned}
\end{equation}

\subsubsection{Second Order}

\begin{figure}[t]
  			\centering
    		\includegraphics[width=0.25\textwidth]{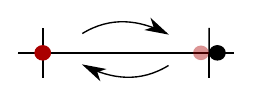}
    		\caption{Second order contributions to the effective Hamiltonian can be viewed as a virtual hopping of an auxiliary boson (red ball) to a nearest-neighbor vertex, where other boson lies (black ball), and then hopping back to its original position.}
    		\label{fig:second_order_hopping}
\end{figure}

The second order contributions are given by
\begin{equation}
\label{eq:H_eff_2}
\hat{H}_{\mathrm{eff},\alpha}^{(2)}=-\hat{P}_0 \hat{H}_\mathrm{int}\,\hat{K}\,\hat{H}_\mathrm{int} \hat{P}_0\,,
\end{equation}
where
\begin{equation}
\hat{K}=\sum_{\ket{\phi}\notin \mathcal{H}^0}\frac{\ket{\phi}\bra{\phi}}{E_\phi-E_0}
\end{equation}
projects to states outside the ground state sector. The expression (\ref{eq:H_eff_2}) can be interpreted as second order virtual processes for the auxiliary bosons (Fig. \ref{fig:second_order_hopping}).

Note that, if the operators on the link were unitary, they would cancel with the respective adjoint operators, resulting in a constant contribution. Since this is not the case, however, for a finite number of atoms $N_{0,l}$, the second order virtual processes result in non-trivial corrections to the effective Hamiltonian. We  calculate them explicitly in this case, for clarity. Higher other contributions can be obtained in a similar way.

The operator $\hat{K}$ can be written as
\begin{equation}
\hat{K}=\frac{1}{2\lambda}\sum_{\phi_1}\ket{\phi_1}\bra{\phi_1}\,,
\end{equation}
where states $\ket{\phi_1}$ correspond to a distribution with zero auxiliary bosons at a single vertex, two in a nearest-neighbor one, and one in the rest of them. The factor $2\lambda$ comes from the difference in energy between $\ket{\phi_1}$ and the ground state. The total contribution to (\ref{eq:H_eff_2}) is
\begin{equation}
\begin{aligned}
\hat{H}^{(2)}_{\mathrm{eff}}=&-\frac{\epsilon^2}{2\lambda}\sum_{\phi_1} \sum_{\mathbf{n},k}\sum_{\mathbf{n}^\prime,k^\prime}\ket{\phi_0}\bra{\phi_0}\left(\hat{\chi}^\dagger_{\mathbf{n}\vphantom{\hat{k}}}\hat{\mathcal{U}}^{\dagger}_{\mathbf{n}\vphantom{\hat{k}},k}\hat{\chi}^{\vphantom{\dagger}}_{\mathbf{n}+\hat{k}}+h.c.\right)\ket{\phi_1}\bra{\phi_1}\left(\hat{\chi}^\dagger_{\mathbf{n}^\prime\vphantom{\hat{k}}}\hat{\mathcal{U}}^{\dagger}_{\mathbf{n}^\prime\vphantom{\hat{k}},k^\prime}\hat{\chi}^{\vphantom{\dagger}}_{\mathbf{n}^\prime+\hat{k}^\prime}+h.c.\right)\ket{\phi_0}\bra{\phi_0}\,.
\end{aligned}
\end{equation}
The terms including hopping must connect the states $\ket{\phi_0}$ and $\ket{\phi_1}$ twice. The second order contribution to the effective Hamiltonian reduces, then, to
\begin{equation}
\begin{aligned}
\hat{H}^{(2)}_{\mathrm{eff}}&=-\frac{\epsilon^2}{2\lambda}\sum_{\mathbf{n},k}\left(\hat{\mathcal{U}}^{\vphantom{\dagger}}_{\mathbf{n},k}\hat{\mathcal{U}}^{\dagger}_{\mathbf{n},k}+\hat{\mathcal{U}}^{\dagger}_{\mathbf{n},k}\hat{\mathcal{U}}^{\vphantom{\dagger}}_{\mathbf{n},k}\right)P_0=\frac{\epsilon^2}{\lambda}\frac{4}{N_{0,l}(N_{0,l}+2)}\sum_{\mathbf{n},k} \hat{E}^2_{\mathbf{n},k}P_0\,.
\end{aligned}
\end{equation}
Since this operator is acting on the ground state sector, $P_0$ can be removed.

\subsubsection{Third Order}

There are two types of third order contributions,
\begin{equation}
\label{eq:H_eff_3}
\begin{aligned}
&\hat{H}_{\mathrm{eff}}^{(3)}=\hat{P}_0 \hat{H}_\mathrm{int}\,\hat{K}\,\hat{H}_{\mathrm{eff}}^{(1)}\,\hat{K}\,\hat{H}_\mathrm{int} \hat{P}_0-\frac{1}{2}\{\hat{P}_0 \hat{H}_\mathrm{int}\,\hat{K}^2\,\hat{H}_\mathrm{int} \hat{P}_0,\hat{H}_{\mathrm{eff}}^{(1)}\}\,.
\end{aligned}
\end{equation}
The first one involves, again, the virtual hopping of auxiliary bosons to the nearest-neighbor vertices. The difference is that, in this case, the the Hamiltonian $H_{\mathrm{eff}}^{(1)}$ (which does not change the sector) is applied before the auxiliary boson comes back to its original vertex. This gives rise to non-local contributions, however, they will cancel with the second part of (\ref{eq:H_eff_3}), which is itself non-local. The latter is composed of the first and second order corrections. Performing similar computations as in the second order case, we obtain the following local terms,
\begin{equation}
\begin{aligned}
\hat{H}^{(3)}_{\mathrm{eff}}=&-\frac{\epsilon^2\epsilon^\prime}{\lambda^2}\frac{2}{N^{2}_{0,l}}\sum_{\mathbf{n},k}\Bigg[\frac{1}{N^2_{0,l}}\left(1-2\hat{E}_{\mathbf{n},k}+2\hat{E}^2_{\mathbf{n},k}+\mathcal{O}\left(N^{-1}_{0,l}\right)\right)\hat{\Phi}^\dagger_{\mathbf{n}\vphantom{\hat{k}}}\hat{\mathcal{U}}^{\dagger}_{\mathbf{n}\vphantom{\hat{k}},k}\hat{\Phi}^{\vphantom{\dagger}}_{\mathbf{n}+\hat{k}+\hat{k}}+h.c.\Bigg]\\
&-\frac{\mu\epsilon^2}{\lambda^2}\frac{2}{N_{0,l}(N_{0,l}+2)}\sum_{\mathbf{n}} \hat{E}^2_{\mathbf{n},k}\,.
\end{aligned}
\end{equation}
We get another renormalization of the electric part of the Hamiltonian, as well as a new term that correct the hopping term of dynamical bosons with a function of the electric field on the link. These correction terms disappear in the limit $N_{0,l}\rightarrow\infty$.

\begin{figure}[t]
  			\centering
    		\includegraphics[width=0.7\textwidth]{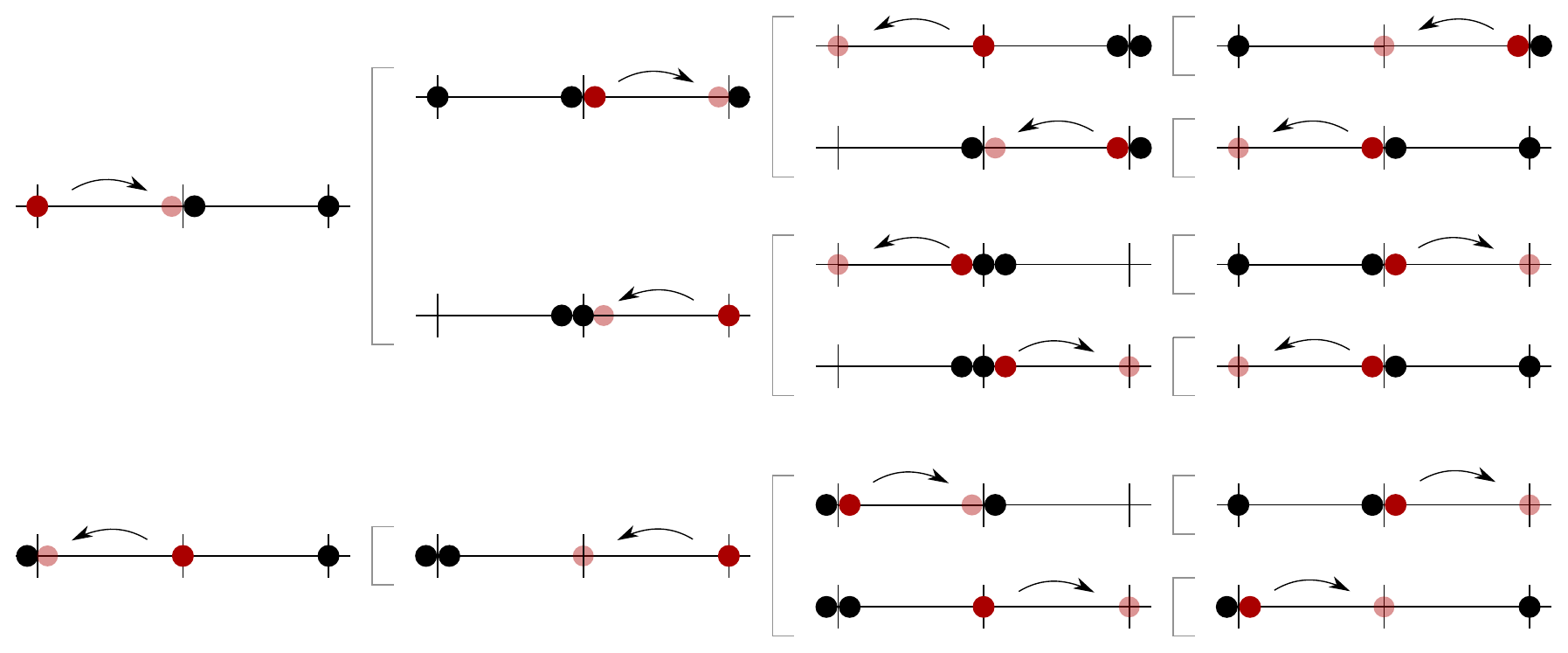}
    		\caption{Many fourth order virtual hopping processes contribute to the effective nearest-neighbor electric field interaction. All the possibilities are represented for two links when the first one is traversed first. The same number of them also appear if one starts from the second link, which are symmetric to the ones represented above. }
    		\label{fig:many_hoppings}
\end{figure}

\subsubsection{Fourth Order}

Many virtual processes contribute to the fourth order corrections to the effective Hamiltonian,
\begin{equation}
\label{eq:H_eff_4}
\begin{aligned}
H_{\mathrm{eff}}^{(4)}=&\frac{1}{2}\{\hat{P}_0 \hat{H}_\mathrm{int}\,\hat{K}\,\hat{H}_{\mathrm{eff}}^{(1)}\,\hat{K}^2\,\hat{H}_\mathrm{int} \hat{P}_0,\,\hat{H}^{(1)}_{\mathrm{eff}}\}+\frac{1}{2}\{\hat{P}_0 \hat{H}_\mathrm{int}\,\hat{K}^2\,\hat{H}_{\mathrm{eff}}^{(1)}\,\hat{K}\,\hat{H}_\mathrm{int} \hat{P}_0,\,\hat{H}^{(1)}_{\mathrm{eff}}\}\\
&+\frac{1}{2}\{\hat{P}_0 \hat{H}_\mathrm{int}\,\hat{K}^2\,\hat{H}_\mathrm{int} \hat{P}_0,\,\hat{P}_0 \hat{H}_\mathrm{int}\,\hat{K}\,\hat{H}_\mathrm{int} \hat{P}_0\}-\frac{1}{2}\{\hat{P}_0 \hat{H}_\mathrm{int}\,\hat{K}^3\,\hat{H}_\mathrm{int} \hat{P}_0,\,\hat{H}^{(1)}_{\mathrm{eff}}\}\\
&-\hat{P}_0 \hat{H}_\mathrm{int}\,\hat{K}\,\hat{H}^{(1)}_{\mathrm{eff}}\,\hat{K}\,\hat{H}^{(1)}_{\mathrm{eff}}\,\hat{K}\,\hat{H}_\mathrm{int} \hat{P}_0-\hat{P}_0 \hat{H}_\mathrm{int}\,\hat{K}\,\hat{H}_\mathrm{int}\,\hat{K}\,\hat{H}_\mathrm{int}\,\hat{K}\,\hat{H}_\mathrm{int} \hat{P}_0\,.
\end{aligned}
\end{equation}

Most of these terms are composed of first, second and third order processes. As in the third order case, many non-local contributions arise, but they all cancel out in the end. Let us consider first the last term in (\ref{eq:H_eff_4}). It involves the four virtual hopping steps of an auxiliary boson, returning to its original position. There are two main processes attached to this. The first one, already explained, related to the hopping  of an auxiliary boson around the fourth links of a plaquette (\ref{fig:hopping_plaquette}), contributing to the plaquette-type interaction,
\begin{equation}
\begin{aligned}
&-\frac{5}{2}\frac{\epsilon^4}{\lambda^3}\sum_{\mathbf{n},ik}\left(\hat{\mathcal{U}}^{\vphantom{\dagger}}_{\mathbf{n}\vphantom{+\hat{k}},i}\hat{\mathcal{U}}^{\vphantom{\dagger}}_{\mathbf{n}+\hat{i},k}\hat{\mathcal{U}}^\dagger_{\mathbf{n}+\hat{k},i}\hat{\mathcal{U}}^\dagger_{\mathbf{n}\vphantom{+\hat{k}},k}+h.c.\right).
\end{aligned}
\end{equation}

The second non-trivial contribution comes from the virtual process that involves hopping twice to a next-nearest-neighbor vertex, and then hopping back to the original position. This can be done in many different ways (Fig. \ref{fig:many_hoppings}). The total contribution, after cancelling out the non-local terms, involve interactions between the electric fields of nearest-neighbor links,
\begin{equation}
\label{eq:n.n.links}
-\frac{\epsilon^4}{\lambda^3}\frac{2}{3}\frac{1}{\left(N_{0,l}(N_{0,l}+2)\right)^2}\sum_{\mathrm{``n.n."}}\left[\hat{E}_{\mathbf{n},k}\hat{E}_{\mathbf{n}^\prime,k^\prime}-11\hat{E}^2_{\mathbf{n},k}\hat{E}^2_{\mathbf{n}^\prime,k^\prime}\right].
\end{equation}

The rest of the terms in (\ref{eq:H_eff_4}) provide many other contributions. Apart from renormalizations to the electric part of the Hamiltonian, we obtain terms that also depend on different powers of the electric field operator,
\begin{equation}
\label{eq:quartic_E}
\begin{aligned}
&\left[\frac{\mu^2\epsilon^2}{\lambda^3}\left(-1+\frac{9}{N_{0,l}(N_{0,l}+2)}\right)+\frac{\epsilon^4}{\lambda^3}\frac{18}{N_{0,l}(N_{0,l}+2)}\right] \sum_{\mathbf{n}} \hat{E}^2_{\mathbf{n},k}\\
&+\frac{\epsilon^2\mu^2}{\lambda^3}\frac{8}{N_{0,l}(N_{0,l}+2)}\sum_{\mathbf{n}} \hat{E}^3_{\mathbf{n},k}+\frac{\epsilon^2\mu^2}{\lambda^3}\frac{6}{N_{0,l}(N_{0,l}+2)}\sum_{\mathbf{n}} \hat{E}_{\mathbf{n},k}\\
&+\frac{\epsilon^4}{\lambda^3}\frac{4}{N_{0,l}(N_{0,l}+2)}\left(\frac{\mu^2}{\epsilon^2}+\frac{2}{N_{0,l}(N_{0,l}+2)}\right)\sum_{\mathbf{n}} \hat{E}^4_{\mathbf{n},k}\,.
\end{aligned}
\end{equation}

We also obtain interactions between the electric field on the links and the dynamical charge operators on the vertices,
\begin{equation}
\begin{aligned}
&\frac{\epsilon^2\epsilon^{\prime\,2}}{\lambda^3}\frac{1}{N^2_{0,l}}\sum_{\mathbf{n},k}\Bigg[\frac{16}{N^2_{0,l}}\hat{E}^2_{\mathbf{n},k}\left(1+\frac{1}{N_{0,v}}(\hat{Q}_{\mathbf{n}\vphantom{+\hat{k}}}+\hat{Q}_{\mathbf{n}+\hat{k}}+1)\right)\\
&+\frac{2}{N^2_{0,v}}\left(\hat{E}_{\mathbf{n},k}(\hat{Q}_{\mathbf{n}\vphantom{+\hat{k}}}-\hat{Q}_{\mathbf{n}+\hat{k}})-\hat{Q}_{\mathbf{n}\vphantom{+\hat{k}}}\hat{Q}_{\mathbf{n}+\hat{k}}+\mathcal{O}\left(N^{-1}_{0,l}\right)\right)\Bigg].
\end{aligned}
\end{equation}

Again, as in the third order, we get corrections to the interaction term between the links and the vertices, that depend, in this case, on the electric field and the charge operators,
\begin{equation}
\begin{aligned}
&\frac{\epsilon^2\epsilon^\prime\mu}{\lambda^3}\frac{1}{2N^{2}_{0,l}}\sum_{\mathbf{n},k}\Bigg[\Bigg((-3+4\hat{E}_{\mathbf{n},k}-4\hat{E}^2_{\mathbf{n},k})+\frac{1}{N_{0,l}}(6-8\hat{E}_{\mathbf{n},k}+8\hat{E}^2_{\mathbf{n},k})\\
&+\frac{1}{N^2_{0,l}}(-4-4\hat{E}_{\mathbf{n},k}+8\hat{E}^2_{\mathbf{n},k})+\mathcal{O}\left(N^{-3}_{0,l}\right)\Bigg)\,\hat{\Phi}^\dagger_{\mathbf{n}\vphantom{\hat{k}}}\hat{\mathcal{U}}^{\dagger}_{\mathbf{n}\vphantom{\hat{k}},k}\hat{\Phi}^{\vphantom{\dagger}}_{\mathbf{n}+\hat{k}}+h.c.\Bigg]
\end{aligned}
\end{equation}
and
\begin{equation}
\begin{aligned}
&\frac{\epsilon^2\epsilon^\prime\mu^\prime}{\lambda^3}\frac{1}{N_{0,l}}\sum_{\mathbf{n},k}\Bigg[\frac{1}{N_{0,l}+2}\left(-1+2\hat{E}_{\mathbf{n},k}\right)\left(\hat{Q}_{\mathbf{n}\vphantom{+\hat{k}}}-\hat{Q}_{\mathbf{n}+\hat{k}}+1\right)\hat{\Phi}^\dagger_{\mathbf{n}\vphantom{\hat{k}}}\hat{\mathcal{U}}^{\dagger}_{\mathbf{n}\vphantom{\hat{k}},k}\hat{\Phi}^{\vphantom{\dagger}}_{\mathbf{n}+\hat{k}}+h.c.\Bigg]\,.
\end{aligned}
\end{equation}

Finally, two new interaction terms appear in the effective Hamiltonian, consisting on applying twice the hopping term. The first one acts on the same link,
\begin{equation}
\begin{aligned}
\frac{\epsilon^2\epsilon^{\prime\,2}}{\lambda^3}\frac{1}{N^2_{0,l}}\sum_{\mathbf{n},k}&\Bigg[\left(1-\frac{2}{N_{0,l}}+\frac{4}{N^2_{0,l}}(2-2\hat{E}_{\mathbf{n},k}+\hat{E}^2_{\mathbf{n},k})+\mathcal{O}\left(N^{-3}_{0,l}\right)\right)\left(\hat{\Phi}^\dagger_{\mathbf{n}\vphantom{\hat{k}}}\hat{\mathcal{U}}^{\dagger}_{\mathbf{n}\vphantom{\hat{k}},k}\hat{\Phi}^{\vphantom{\dagger}}_{\mathbf{n}+\hat{k}}\right)^2+h.c.\Bigg]\,,
\end{aligned}
\end{equation}
whereas the second involve hopping processes in two nearest-neighbor links,
\begin{equation}
\begin{aligned}
&\frac{\epsilon^2\epsilon^{\prime\,2}}{\lambda^3}\frac{1}{2N_{0,l}(N_{0,l}+2)}\sum_{\mathrm{``n.n."}}\Bigg[\left(1-2\hat{E}_{\mathbf{n},k}\right)\left[\left(\hat{\Phi}^\dagger_{\mathbf{n}\vphantom{\hat{k}}}\hat{\mathcal{U}}^{\dagger}_{\mathbf{n}\vphantom{\hat{k}},k}\hat{\Phi}^{\vphantom{\dagger}}_{\mathbf{n}+\hat{k}}+h.c.\right),\hat{\Phi}^\dagger_{\mathbf{n}\vphantom{\hat{k}}}\hat{\mathcal{U}}^{\dagger}_{\mathbf{n}\vphantom{\hat{k}},k}\hat{\Phi}^{\vphantom{\dagger}}_{\mathbf{n}+\hat{k}}\Bigg]+h.c\right],
\end{aligned}
\end{equation}
and it is written in terms of the commutator between them.

\subsection{Relevant Corrections}
\label{app:negligible}

We study the importance of the correction terms (for the case $\mu=0$), obtained in the effective expansion of the Hamiltonian, by expressing them, first, in terms of the simulated parameters, $R$ (\ref{eq:R_g}) and $g$ (\ref{eq:g}), and multiplying everything by $\alpha$ (\ref{eq:alpha}). Let us consider the terms proportional to $\left(\epsilon/\lambda\right)^4$. They are proportional to the following coupling constants,

\begin{equation}
-\frac{\epsilon^4}{\lambda^3}\frac{2}{3}\frac{1}{\left(N_{0,l}(N_{0,l}+2)\right)^2}\alpha=-\frac{2}{15g^2}\frac{1}{\left(N_{0,l}(N_{0,l}+2)\right)^2}\,,
\end{equation}

\begin{equation}
\frac{\epsilon^4}{\lambda^3}\frac{8}{\left(N_{0,l}(N_{0,l}+2)\right)^2}\alpha=\frac{8}{5g^2}\frac{1}{\left(N_{0,l}(N_{0,l}+2)\right)^2}\,,
\end{equation}
which correspond to the electric field interaction between nearest-neighbor links (\ref{eq:n.n.links}) and the quartic electric field term (\ref{eq:quartic_E}), respectively. These terms become relevant in the weak regime. However, their importance decreases rapidly when the number of atoms increases.

The rest of the non-desired contributions are proportional to the following parameters,
\begin{equation}
\frac{\epsilon^2\epsilon^{\prime}}{\lambda^2}\frac{1}{N^2_{0,l}}\alpha=\frac{R^2}{N^2_{0,l}}\left(\frac{\epsilon}{\lambda}\right)^2\ll 1\,,
\end{equation}

\begin{equation}
\frac{\epsilon^2\epsilon^{\prime\,2}}{\lambda^3}\frac{1}{N^2_{0,l}}\alpha=\frac{5R^4g^2}{N^2_{0,l}}\left(\frac{\epsilon}{\lambda}\right)^6\ll 1\,,
\end{equation}

\begin{equation}
\frac{\epsilon^2\epsilon^{\prime}\mu^\prime}{\lambda^3}\frac{1}{N_{0,l}}\alpha=-\frac{5}{4}\frac{g^2}{N_{0,l}}\left(\frac{\epsilon}{\lambda}\right)^6\ll 1\,.
\end{equation}
All of them are negligible, since $\frac{\epsilon}{\lambda}$ can be made small enough to compensate for the possible growth due to $R$ and $g$. We denote all the terms of this type with $\hat{H}^\prime_\mathrm{eff}$. In general, we can say that only the corrections of order $\left(\epsilon/\lambda\right)^4$ are important, since this is the order of the effective plaquettes, and is the most relevant one in the weak regime.

\section*{References}

\bibliography{ref}

\end{document}